\documentclass[12pt,preprint]{aastex}
\usepackage{emulateapj5,apjfonts,psfig}

\newcommand{\etal}{{et~al.}}

\newcommand{\Msun}{M_\odot}

\newcommand{\Lsun}{L_\odot}
\newcommand{\kms}{$\rm {km}~\rm s^{-1}$}
\newcommand{\mtl}{{\it M/L}}

\begin{document}

\lefthead{Galaxy Models}
\righthead{Gebhardt~\etal}

\title{Axisymmetric Dynamical Models of the Central Regions of Galaxies}

\author{Karl Gebhardt\altaffilmark{1}, Douglas
Richstone\altaffilmark{2}, Scott Tremaine\altaffilmark{3}, Tod
R. Lauer\altaffilmark{4}, Ralf Bender\altaffilmark{5}, Gary
Bower\altaffilmark{6}, Alan Dressler\altaffilmark{7},
S.M.~Faber\altaffilmark{8}, Alexei V. Filippenko\altaffilmark{9},
Richard Green\altaffilmark{4}, Carl Grillmair\altaffilmark{10}, Luis
C.  Ho\altaffilmark{7}, John Kormendy\altaffilmark{1}, John
Magorrian\altaffilmark{11}, and Jason Pinkney\altaffilmark{2}}

\altaffiltext{1}{Department of Astronomy, University of Texas, Austin,
Texas 78712; gebhardt@astro.as.utexas.edu, kormendy@astro.as.utexas.edu}
 
\altaffiltext{2}{Dept. of Astronomy, Dennison Bldg., Univ. of
Michigan, Ann Arbor 48109; dor@astro.lsa.umich.edu,
jpinkney@astro.lsa.umich.edu}
 
\altaffiltext{3}{Princeton University Observatory, Peyton Hall,
Princeton, NJ 08544; tremaine@astro.princeton.edu}
 
\altaffiltext{4}{National Optical Astronomy Observatories, P. O. Box
26732, Tucson, AZ 85726; green@noao.edu, lauer@noao.edu}

\altaffiltext{5}{Universit\"ats-Sternwarte, Scheinerstrasse 1,
M\"unchen 81679, Germany; bender@usm.uni-muenchen.de}

\altaffiltext{6}{Computer Sciences Corporation, Space Telescope
Science Institute, 3700 San Martin Drive, Baltimore, MD 21218;
bower@stsci.edu}

\altaffiltext{7}{The Observatories of the Carnegie Institution of
Washington, 813 Santa Barbara St., Pasadena, CA 91101;
dressler@ociw.edu, lho@ociw.edu}
 
\altaffiltext{8}{UCO/Lick Observatories, University of California,
Santa Cruz, CA 95064; faber@ucolick.org}

\altaffiltext{9}{Department of Astronomy, University of California,
Berkeley, CA 94720-3411; alex@astro.berkeley.edu}

\altaffiltext{10}{SIRTF Science Center, 770 South Wilson Ave.,
Pasadena, CA 91125; carl@ipac.caltech.edu}

\altaffiltext{11}{Department of Physics, University of Durham,
Rochester Building, Science Laboratories, South Road, Durham DH1 3LE,
UK; John.Magorrian@durham.ac.uk}

\begin{abstract}

We present axisymmetric, orbit superposition models for 12 galaxies
using data taken with the {\it Hubble Space Telescope (HST)} and
ground-based observatories. In each galaxy, we detect a central black
hole (BH) and measure its mass to accuracies ranging from 10\% to
70\%. We demonstrate that in most cases the BH detection requires {\it
both} the {\it HST} and ground-based data.  Using the ground-based
data alone does provide an unbiased measure of the BH mass (provided
they are fit with fully general models), but at a greatly reduced
significance. The most significant correlation with host galaxy
properties is the relation between the BH mass and the velocity
dispersion of the host galaxy; we find no other equally strong
correlation, and no second parameter that improves the quality of the
mass-dispersion relation. We are also able to measure the stellar
orbital properties from these general models. The most massive
galaxies are strongly biased to tangential orbits near the BH,
consistent with binary BH models, while lower-mass galaxies have a
range of anisotropies, consistent with an adiabatic growth of the BH.

\end{abstract}
 
\keywords{galaxies: nuclei --- galaxies: statistics --- galaxies: general}

\section{Introduction}

Most nearby galaxies contain massive compact dark objects at their
centers. The number density and masses of these objects are consistent
with the hypothesis that they are dead quasars: massive black holes
that grew mainly by gas accretion and were once visible as quasars or
other active galactic nuclei from radiation emitted during the
accretion process (see Kormendy~\& Richstone 1995 for a review).

We have obtained {\it Hubble Space Telescope (HST)} spectra of
the centers of 12 nearby galaxies, using first the square aperture of
the Faint Object Spectrograph (FOS) and later the long-slit on the
Space Telescope Imaging Spectrograph (STIS). Additional ground-based
spectra have been obtained at the MDM Observatory. Pinkney~\etal\
(2002a) describe the data collected by our group for the 10 galaxies
observed with STIS, and we present the data for the two galaxies
observed with FOS in the Appendix of this paper. Section~2 discusses
how we incorporate the data into the dynamical models.

An overall discussion of the dynamical modeling methods is given in
Gebhardt~\etal\ (2000a) and Richstone~\etal\ (2002). The models are
axisymmetric and based on superposition of individual stellar
orbits. Section~3 provides the details of the models for these
galaxies. Five other galaxies have stellar-dynamical data and models
of comparable quality. Three of these are from the Leiden group: M32
(van~der~Marel~\etal\ 1998; Verolme~\etal\ 2002), NGC~4342 (Cretton~\&
van den Bosch 1999), and IC~1459 (Cappellari~\etal\ 2002). The
remaining two are NGC~3379 (Gebhardt~\etal\ 2000a) and NGC~1023
(Bower~\etal\ 2001). Results from these five additional galaxies are
included in the analysis in Section~4.


\begin{figure*}[b]
\centerline{\psfig{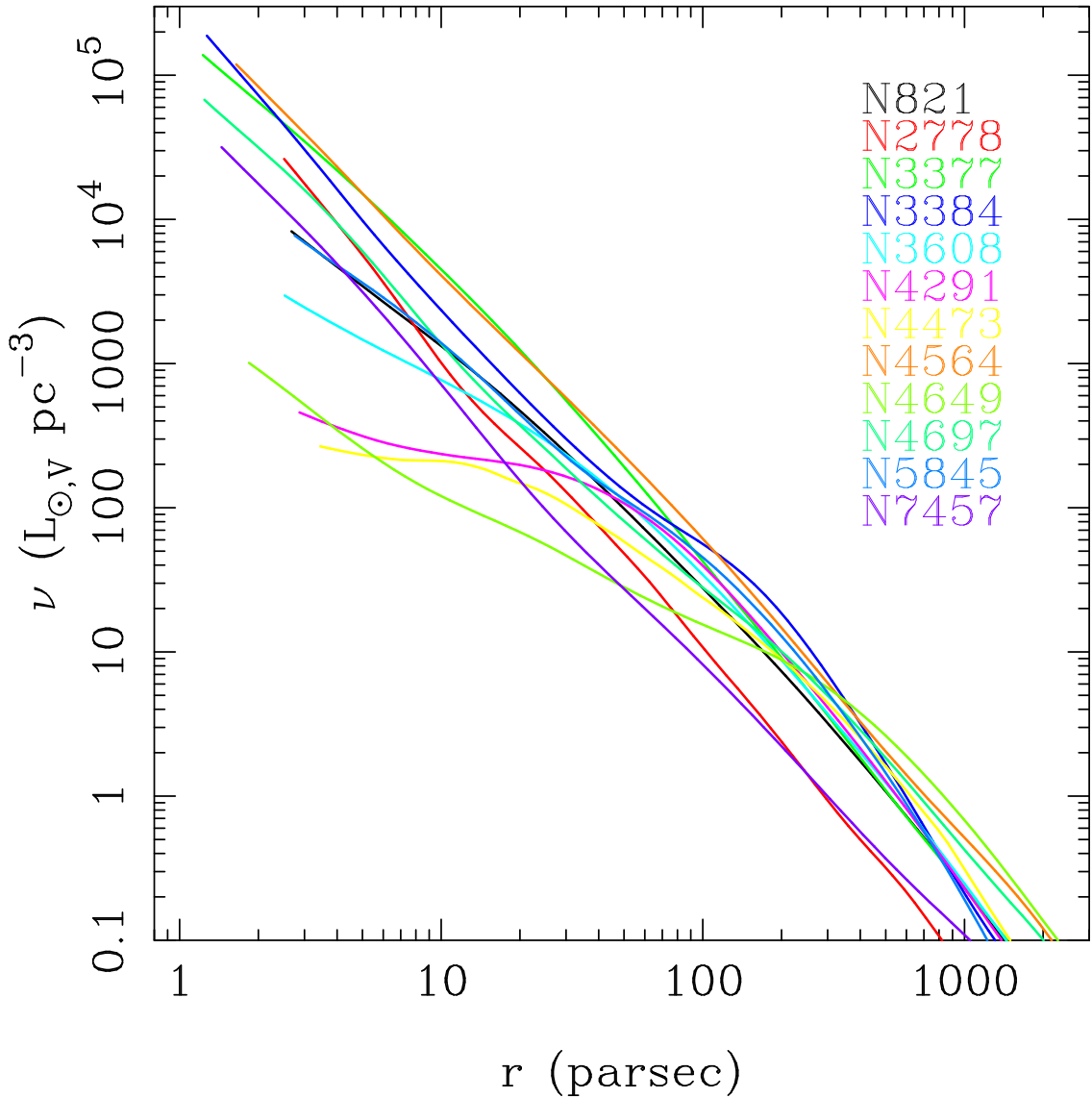}}
\vskip 0pt \figcaption[gebhardt.fig1.ps]{Luminosity density profiles
for the sample galaxies. These are in the $V$ band and include both {\it
HST} and ground-based data. The radii are along the semi-major
axis.
\label{fig1}}
\end{figure*}


We use orbit-based models rather than parameterized models of the
distribution function because parameterization can lead to biased
black hole (BH) mass estimates. Parameterized models can even imply the
presence of a BH when none exists. Orbit-based models do not
suffer from this bias. However, we do make various assumptions whose
consequences must be examined (Section~5). In particular, we model
galaxies as axisymmetric. Triaxial, and worse yet, asymmetric
galaxies, may be poorly represented by axisymmetric models. However,
these effects are likely to be random and therefore it is reasonable
to expect that the assumption of axisymmetry will not cause an overall
bias in the BH mass.

In addition to measuring the BH mass ($M_{BH}$) and stellar
mass-to-light ratio (\mtl, assumed to be independent of position), our
models constrain the orbital structure in the galaxy. It appears from
this study and those of Verolme~\etal\ (2002) and Cappellari~\etal\
(2002) that the distribution function in axisymmetric galaxies depends
on all three integrals of motion, not just the energy and angular
momentum.

Preliminary BH masses for these galaxies have been reported by
Gebhardt~\etal\ (2000b); these masses are based on a coarser grid of
models (explained in Section~4) and thus have larger uncertainties
than those presented here. However, the best-fit values for the 
BH masses are nearly the same in the two studies.

Most distances in this paper have been measured with the
surface-brightness fluctuation method (SBF, Tonry~\etal\ 2000); for
those galaxies without an SBF distance we assume the distance in an
unperturbed Hubble flow and $H_0 = 80$ km s$^{-1}$ Mpc$^{-1}$.

\section{Data}

The data consist of images and spectra from ground-based and {\it HST}
observations. The high spatial resolution of {\it HST} is essential to
measure the mass of the central BH. The ground-based data are
essential to constrain the stellar orbital distribution and
mass-to-light ratio. Since we are using two-dimensional galaxy models,
we must have data along various position angles to constrain
adequately the orbital structure.

\subsection{Imaging}

Most of the sample galaxies were imaged with WFPC2 during {\it HST}
Cycles 4 and 5; the exception is NGC 4697, which was observed with
WFPC1 (see Lauer~\etal\ 1995).  In general, each galaxy was observed
in both the F555W ($V$) and F814W ($I$) filters. The typical total
integration time in each filter was $\sim1200$~s, but subdivided into
shorter exposures to allow for the identification of cosmic-ray
events.  The exposure levels at the centers of all galaxies exceeded
$10^4$ photons pixel$^{-1}$, and were often nearly an order of
magnitude higher.  After the sub-exposures were compared to detect and
eliminate cosmic-ray events, they were combined and then deconvolved
using Lucy (1974) and Richardson (1972) deconvolution.  The
point-spread functions (PSFs) were provided by standard-star
observations obtained during the routine photometric monitoring of
WFPC2.  Typically, 40 iterations of Lucy-Richardson deconvolution were
used.  Lauer~\etal\ (1998) demonstrate that this procedure allows
accurate recovery of the intrinsic galaxy brightness distribution for
all but the central pixel.  Brightness profiles were then measured
from the deconvolved images using the high-resolution Fourier
isophote-fitting program of Lauer (1985).  The present work uses the
$V$-band profiles, given their intrinsically higher spatial
resolution. The {\it HST} imaging provides adequate coverage out
to around 10\arcsec; beyond that we rely on ground-based imaging to
complete the radial coverage. Ground-based imaging comes primarily
from Peletier~\etal\ (1990).

Figure~1 presents the luminosity density profiles of the galaxies,
which are input to the dynamical models. We determine the luminosity
density distribution from the surface-brightness distribution by
assuming that the luminosity density is axisymmetric, and constant on
similar spheroids. (In one galaxy, NGC 4473, we have included a
stellar disk in addition to the spheroidal luminosity distribution.)
We use the non-parametric techniques described by Gebhardt~\etal\
(1996), which involve smoothing the surface brightness and then
inverting the Abel integral equation that relates surface brightness
and luminosity density. We note that without this or some other
restrictive assumption on the shape of the equidensity surfaces, the
deprojection is not unique, except for edge-on galaxies (Gerhard \&
Binney 1996, Kochanek \& Rybicki 1996). In particular, Magorrian \&
Ballantyne (2001) show that deprojection uncertainties, and in
particular face-on disks, can significantly increase the uncertainties
in the measured orbital distribution. We do not attempt a complete
treatment of deprojection uncertainties in this paper, but do discuss
possible consequences and biases in Section 4.9.

\subsection{HST Kinematics}

Pinkney~\etal\ (2002a) present the spectra and kinematics from the
{\it HST} STIS observations. Most of the galaxies in our sample have
spectra taken with STIS, except for NGC~3377 and NGC~5845, which were
observed with a single FOS aperture. The kinematic results for these
two galaxies are presented in the Appendix. We use the line-of-sight
velocity distributions (LOSVDs) in the modeling (i.e., we fit to the
binned LOSVD, not its moments). For most of the galaxies, we use 13
equally-spaced velocity bins to represent the LOSVD. The width of the
velocity bins is generally around 40\% of the galaxy's velocity
dispersion. The uncertainty in the signal in each velocity bin is
determined using Monte Carlo simulations. We reproduce a sample of the
velocity profiles in Figure~2 (for NGC~4564).

The LOSVD can be biased by several systematic effects, including the
choice of stellar template, continuum shape, spectral range used in
the fit, and amount of smoothing. These are discussed by
Pinkney~\etal\ (2002a). In general, the most significant bias is
probably template mismatch. However, most of our data are observed in
the Ca~II near-triplet region (8500~\AA), and in this region the LOSVD
is not very sensitive to template variations.

There is scattered light in STIS that is about 0.2\% of the incoming
light. The scattering occurs after the light has passed through the
grating and so is not due to the PSF of {\it HST}. We measure this
light using the spectral lamp images where we can use the high
signal-to-noise ratio (S/N) to study the wings of the profile. There
is a broad component that has a standard deviation equal to 25 pixels,
presumably due to scattering in the STIS optics. We have run extensive
tests to determine whether this scattered light affects our
results. It is possible that a bright nucleus can scatter light into
neighboring pixels which would not be reflected in the assumed PSF. We
simulate this effect in both NGC~3377 (a power-law galaxy) and M87 (a
core galaxy). We simulate the two-dimensional image by inputing
kinematic profiles consistent with those measured in both galaxies,
and then convolving those kinematics with both the narrow and broad
components of the PSF. We then extract and fit the profiles ignoring
the broad component, and compare with the input values. There was
essentially no effect in M87, as expected since its kinematic profile
does not vary strongly with radius. For NGC~3377, the change in the
second moment was negligible; however, the velocity profile showed a
5\% reduction in peak amplitude and the dispersion profile showed a
5\% increase near the center. Since the second moment was hardly
changed, this broad component has no effect on the measured BH mass
and is not included in subsequent analysis.

The STIS spectral resolution with the G750M grating is around 55~\kms\
(FWHM), with 37~\kms\ per binned pixel (we bin 2x1 pixels on the chip
for most of our data). Since we always use a template star convolved
with the LOSVD to match the galaxy spectrum, we do not have to worry
as much about the detailed shape of the spectral PSF (as opposed to
understanding the spatial PSF) because the velocity dispersions of our
galaxies are much larger than the spectral FWHM. The main concern is
whether we are illuminating the slit with the templates in the same
way that we illuminate with the galaxy. For galaxies with point-like
nuclei, this is not a concern, but for those with shallow light
profiles we must consider the effect. The concern is whether the
velocity variation across the slit adds to the dispersion measured in
the galaxy, which would not be true for a point source. We can
calculate the effect using the results of Bower~\etal\ (2001). For the
0.1\arcsec\ slit, the velocity variation from slit edge to edge is
around 20~\kms, and for the 0.2\arcsec\ slit, it is 40~\kms. Given the
FWHM of the spectral lines, 55~\kms, the velocity variation provides a
7--25\% increase in the required instrumental spectral FWHM. However,
for the galaxies where this effect is the largest (i.e., the core
galaxies), the galaxy dispersion is the greatest and therefore the
broadening in the galaxy is insignificantly affected by the exact
instrumental profile. For example, in the case of NGC~3608, the
central dispersion is 300~\kms; with a change in the instrumental
dispersion from 20~\kms\ to 25~\kms\ (25\% higher), the inferred
galaxy dispersion differs by only 0.1\%. Given the insignificant
difference, we apply no correction for illumination
effects. Bower~\etal\ (2001) find a similar result for NGC~1023.

\subsection{Ground-Based Kinematics}

Nearly all of the ground-based data come from the MDM Observatory
(Pinkney~\etal\ 2002a). Briefly, most of the spectra were taken around
the Ca~II triplet (near 8500~\AA) and the rest were taken near the Mg
$b$ region (5100~\AA). The instrumental resolution varied slightly
from run to run, but was generally around 40~\kms, which is more than
adequate given the dispersions of the galaxies studied. The spatial
resolution varied from 0.5\arcsec\ to 1.5\arcsec. We included the
appropriate spatial PSF for each of the ground-based spectra in the
modeling.

In an axisymmetric system, the velocity profile at a radius on one
side of the galaxy will be identical to a profile that is flipped
about zero velocity on the other side of the galaxy at the same
radius. There are three options that we can use to include this
symmetry in the models. First, we can fit the same, but appropriately
flipped, observed velocity profile on the two spectra from opposite
sides of the galaxy during the extraction. In this way, we only
include one profile at a given radius. Second, we can independently
fit velocity profiles from opposite sides of the galaxy, and then
average these two velocity profiles (after flipping one of them) to
provide one profile for that radius. Third, we can include the two
independently fit velocity profiles directly into the models. Each of
these has their own advantages and disadvantages. For example, if
there is a bad spot on the detector or a star on one side of the
galaxy, then the most reliable measure would be to use the third
option (since one can then exclude the affected region).

We have tried all three methods and find little differences between
the results.  We choose to use the first option since, in that case,
the S/N used for the extraction of the velocity profile is increased
by $\sqrt2~$ compared to the other cases, and this serves to alleviate
potential biases. This increase arises because we use two spectra to
measure one velocity profile, as opposed to measuring two independent
velocity profiles. The uncertainty (and hence, the S/N) in the
resultant velocity profile is the same regardless of the method used
to estimate it, but our reason for using the first option is motivated
by alleviating potential biases in the extraction of the velocity
profile.  For low S/N data, there are often biases in the velocity
profile (mainly due to the need to use more smoothing as the signal is
lowered), and we decrease these biases by forcing axisymmetry during
the spectral extraction. The alternative of using individual profiles
from both sides of the galaxy is not optimal.


\begin{figure*}[t]
\centerline{\psfig{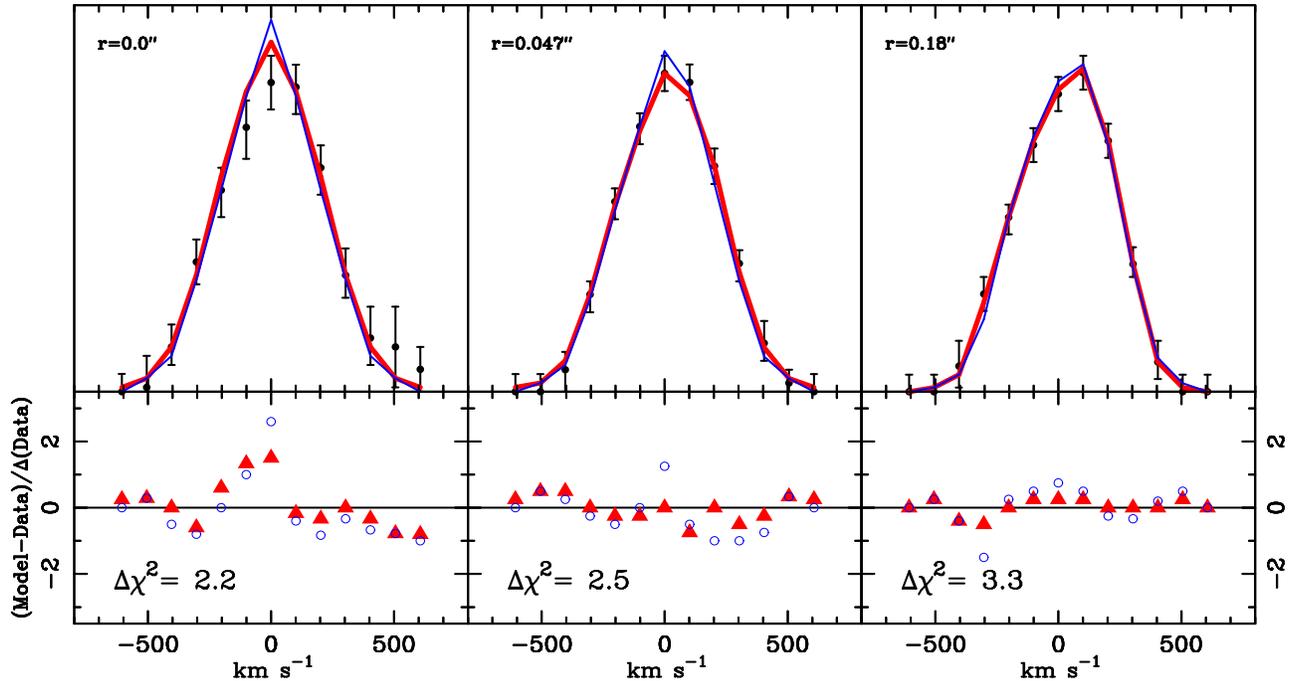}}
\vskip 0pt \figcaption[gebhardt.fig2.ps]{The comparison of the LOSVD
from three of the 33 velocity profiles in NGC~4564; the radii are
given in the upper left for each panel. In the upper panels, the solid
circles with error bars represent the data LOSVDs. The lines in the
upper panel represent two different models, and the lower panel shows
their residuals from the data normalized by dividing by the
uncertainty in the data. The red line and triangles are the values
from the best-fit BH model; the blue line and open circles are the
model without a BH.  The $\Delta\chi^2$ value given at the bottom is
the difference between the best fit model and the zero BH mass model
for that particular bin. These three bins contribute a total of
$\Delta\chi^2=8.0$, whereas the difference from the full sample is
52.7.
\label{fig2}}
\end{figure*}


\section{Dynamical Models}

Richstone~\etal\ (2002) provide a complete account of the construction
of the dynamical models, including analytic tests. Here, we provide a
basic summary and include the details that are specific to these
galaxy models. Other groups discuss the use of and tests for similar
orbit-based models (van der Marel~\etal\ 1998, Cretton~\& van den
Bosch 1999, Cretton~\etal\ 1999, 2000, Cappellari~\etal\ 2002,
Verolme~\& de Zeeuw 2002, and Verolme~\etal\ 2002). These studies used
models similar to each other; the models presented here and in
Gebhardt~\etal\ (2000a) differ from those above in small but important
ways. As discussed in Richstone~\etal\ our models use a maximum
likelihood approach to find the orbital weights as opposed to using a
regularization method, and ours also use the full LOSVD as opposed to
using parameterized moments. There are positives and negatives
associated with the different approaches and a full comparison can
only be studied when the different models are applied to identical
datasets.

The dynamical models are constructed as follows: we first determine
the luminosity density from the surface brightness profile. Although
we have constructed models with a variety of inclinations, we
generally assume that the galaxy is edge-on, for reasons given in
Section 4. In this case, the deprojection is unique. To determine the
potential we assume a stellar mass-to-light ratio and a BH mass. In
this potential we run a representative set of orbits (typically 7000)
that cover phase space adequately. We then find the non-negative set
of weights for those orbits that provides the best match to the
available data (in the sense of minimum $\chi^2$). In order to have a
smooth phase space distribution, we use a maximum entropy method as
described below. We repeat this analysis for different BH masses and
different mass-to-light ratios to find the overall best fit.

We measure the velocity moments of our models on a two-dimensional
grid in radius and angle relative to the symmetry axis of the
galaxy. We generally use 20 radial and 5 angular grid elements. The
parameters of this grid (spacing and extent) are designed to maximize
the S/N in both the kinematics and the photometry. The angular bins
have centers at latitudes 5.8\degr, 17.6\degr, 30.2\degr, 45.0\degr,
and 71.6\degr, where the angle is defined from the major to the minor
axis; we use the same binning scheme whether we are in projected or
internal space. We have run tests in which we both double and halve
the number of bins and we find insignificant differences. Since STIS
provides kinematic information along a slit, we need to specify how to
extract the data along that slit to optimize the S/N to measure the BH
mass. Pinkney~\etal\ (2002a) describe the 20 radial extraction windows
that we use. We define our radial binning scheme in the models with
the same configuration as that used in the data extraction.

We specify the galaxy potential and the forces on a grid that is five
times finer than the grids used in the data comparisons, in order to
assure accurate orbit integration. If we have $N$ radial bins, labeled
$i=1,\ldots,N$, our goal is to have at least one orbit with apocenter
and pericenter in every possible pair of bins $(i,j)$ in order to
cover phase space well; this requires $N(N-1)/2$ orbits, times two to
include stars with the opposite sign of rotation. This leads to 380
orbits; however, we must also cover the angular dependence and to do
this we include 20 additional angular bins. Thus, the total number of
orbits is around 7000. We track the velocity information by storing
the LOSVD for each orbit in each grid element. For each galaxy we use
13 velocity bins, spanning the maximum and minimum velocities
generated for the whole orbit distribution. It is important to include
all velocity information, particularly in the LOSVD wings where the
effects from the BH are the strongest. Our final models consist of
7000 (orbits) x 20 (radial) x 5 (angular) x 13 (velocity)
elements. For each galaxy, we generally try about 10 different BH
masses, 10 or more values of \mtl, and sometimes a few different
inclinations.  We have also run models where we have both doubled and
halved the number of orbits. In either case, we find no difference in
the best fit to the data.

In addition to the projected quantities, we track the internal
properties including the velocity moments and luminosity density. For
the dynamics, we only track the zeroth, first, and second moments of
the velocity profile. The internal moments are presented in Section
4.7 below.

It is important to include the effects of the PSF of {\it HST} in the
dynamical models. We use the same PSF as measured by Bower~\etal\
(2001), which has FWHM = 0.08\arcsec\ along the slit at 8500~\AA. At
8500~\AA, the first diffraction peak is visible and is included in the
PSF model. This profile comes from a highly sampled PSF using a cut in
the spatial direction for a star at various columns on the chip. Since
there is a 6-row shift of the star across the STIS chip, the PSF is
sampled differently in each column, thereby producing a well-sampled
profile. Unfortunately, this procedure only produces a PSF in one
dimension (along the slit), and we have no measurement of the PSF in
the spectral direction. As discussed by Bower~\etal\ (2001), the PSF
is expected to be circularly symmetric, so we assume that the PSF
across the slit is the same as the PSF measured along the slit. We run
all orbit libraries with no PSF included and then convolve with the
appropriate PSF before we fit to the kinematic data. In this way, we
can include a different PSF for each kinematic observation, if
necessary.

We have ground-based kinematic data along 2--4 position angles for
each galaxy, covering over half of the radial bins. Thus, we typically
have 20 positions on the sky, each with 13 LOSVD bins, for a total of
260 data bins. However, as explained by Gebhardt~\etal\ (2000a) and
discussed further in \S4.1, the number of degrees of freedom is
difficult to estimate. The main problem is that the smoothing in the
LOSVD estimation introduces covariance between velocity bins. For a
typical galaxy, every two velocity bins are correlated. This factor of
two is determined through simulations where we vary the smoothing
parameter in the velocity profile estimate and measure the effect on
$\chi^2$ (see Gebhardt~\etal\ 2000a). Thus, the number of degrees of
freedom is reduced by a factor of roughly two compared to the total
observed parameters.

The orbit weights are chosen so that the luminosity density in
every spatial bin matches the observations to better than
1\%. Typically, the match is better than 0.1\%. We regard matching the
luminosity density in each bin as a set of constraints, rather than a
set of data points. Thus, the photometric data do not contribute to
the total number of degrees of freedom. We make this choice for two
reasons: first, the uncertainties in the photometry are much smaller
than those in the kinematics; second, including photometric
uncertainties would require compiling a far larger set of orbit
libraries (one for each tested photometric profile).

To ensure that the phase-space distribution function is smooth, we
maximize the entropy as in Richstone \& Tremaine (1984). We do this by
defining a function $f\equiv \chi^2-\alpha S$, where $\chi^2$ is the
sum of squared residuals to the data, $S$ is the entropy, and $\alpha$
is a parameter describing the relative weights of entropy and
residuals in the fit. Our goal is to minimize $f$. We start with a
large value of $\alpha$ and then gradually reduce it until further
improvement in $\chi^2$ is no longer possible. At first, the entropy
determines the orbital weights but, at the end of the minimization,
the entropy has no influence on the quality of the fit. The entropy
constraint does affect those regions where we do not have kinematic
data, but we never use results from those regions. Solving for the
7000 orbital weights with 200--500 observations is the most
computationally expensive part of the analysis. We have tried a
variety of initial conditions for the orbital weights and entropy
forms, and find that neither the minimum value of $\chi^2$, nor the BH
mass and stellar mass-to-light ratio, nor the orbital structure is
sensitive to these choices.


\begin{figure*}[t]
\centerline{\psfig{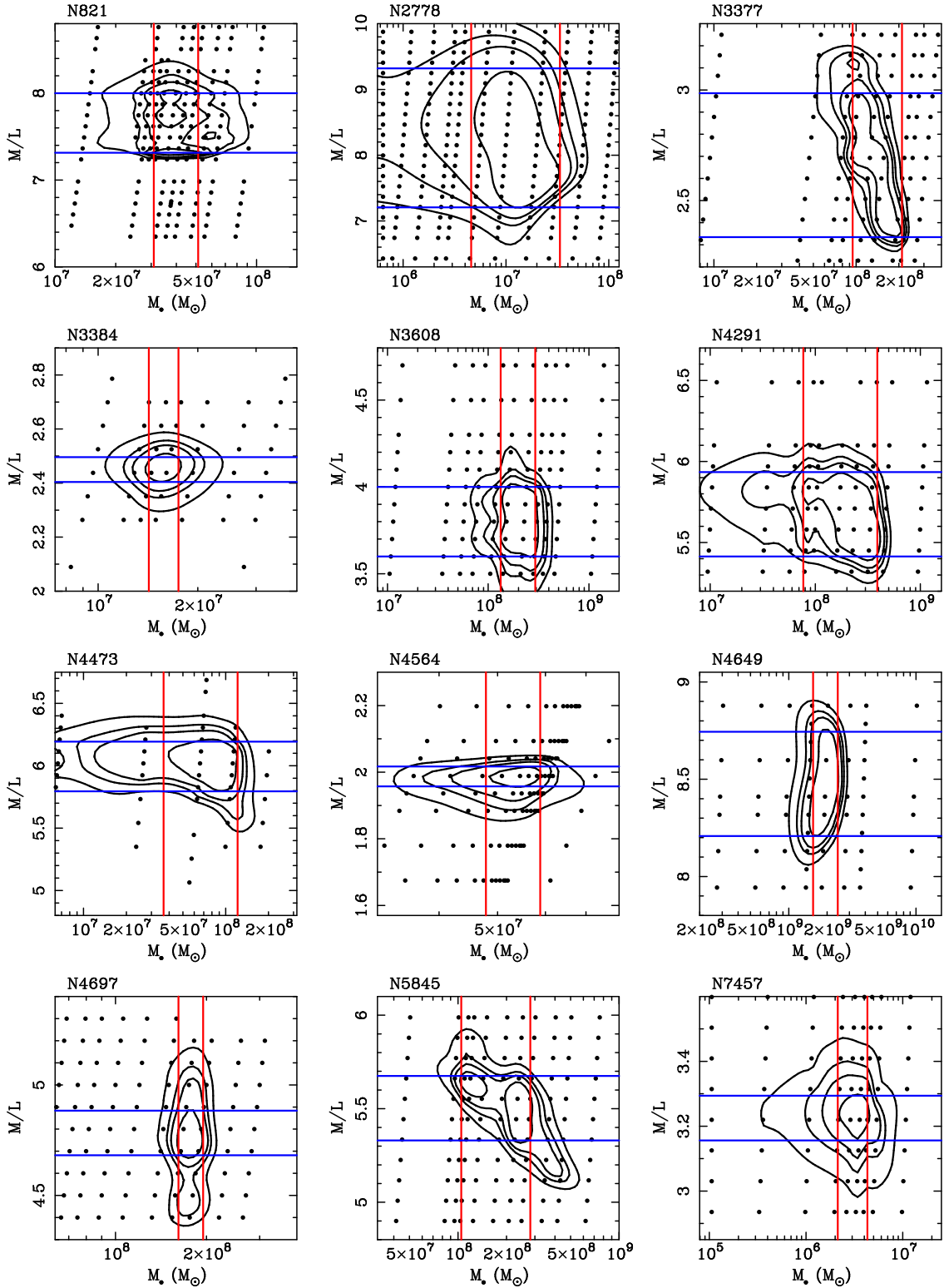}}
\vskip 0pt \figcaption[gebhardt.fig3.ps]{Two-dimensional plots of
$\chi^2$ as a function of BH mass and mass-to-light ratio for each of
the galaxies. The points represent models that we ran. The contours
were determined by a two-dimensional smoothing spline interpolated
from these models, and represent $\Delta\chi^2$ of 1.0, 2.71, 4.0, and
6.63 (corresponding to 68\%, 90\%, 95\%, 99\% for one degree of
freedom). The red vertical lines are the 68\% limits for the BH masses
marginalized over mass-to-light ratio, and the blue horizontal lines
are the 68\% limits for the mass-to-light ratios marginalized over BH
mass.
\label{fig3}}
\end{figure*}


We need to determine the uncertainties in the BH mass and the
stellar \mtl. These are correlated, of course, and we
generally use two-dimensional $\chi^2$ distributions to determine the
uncertainties.  The uncertainties in the parameters are determined
from the change in $\chi^2$ as we vary one of the variables; in this
case, the 68\% confidence band is reached when $\chi^2$ increases
above its minimum value by 1. This parameter estimation is different
from hypothesis testing: a tested hypothesis is consistent with the
data if $\chi^2$ per degree of freedom is approximately unity, while
the allowed range of a parameter is determined by the change in
$\chi^2$ from its minimum value. For example, since the BH has
no effect on the kinematics at large radii, we could always ensure that
a galaxy is consistent with the hypothesis that there is no BH
by adding more and more kinematic data at large radii.

Thus, we advocate that one must use $\Delta\chi^2$ in order to
determine the uncertainties in the parameters. This conclusion was
also discussed in both van der Marel~\etal\ (1998) and Cretton~\etal\
(2000). The difficulty for the parameter estimation is that we need to
measure the uncertainties in the kinematics accurately. The
uncertainties on the kinematics are difficult to quantify; problems
due to template mismatch and continuum estimation, for example, can
have a significant effect on the results.  We have tried to take this
into account during the Monte Carlo simulations that we use to
generate the errors. A more natural approach would be to use a
Bayesian analysis, but given the large number of unknown variables
(the 7000 orbital weights), this is impractical.

Figure~2 shows the data/model comparison for three LOSVDs in NGC~4564.
For this galaxy, we actually have 33 velocity profiles but only show
three here. Thus, there is significantly more data that has gone into
the models. For NGC~4564, the signature for the BH comes from the
central few bins in each of the three position angles. However, it is
only by examining the full dataset can one understand the global fit
for any particular model. The change in $\chi^2$ is given in the
bottom plot. The difference between the best-fit BH and zero BH model
in just these three bins is equal to 8. Using the full dataset the
difference is 53, implying an extremely high significance against the
zero BH model.

\section{Results}

The three main properties which we obtain from the models are the BH
mass, mass-to-light ratio, and the orbital structure. The BH mass and
mass-to-light ratio are fitted by choosing a grid of parameters for
them and then examining their $\chi^2$ distribution. The orbital
structure, however, results from finding the orbital weights for each
specified potential (i.e., BH mass and mass-to-light ratio) that
provides the minimum $\chi^2$. Each BH mass/mass-to-light ratio pair
produces a best fit orbital structure, but the overall best model is
that which has the one global minimum. In this section, we discuss
results for each parameter, and consider possible biases and additional
uncertainties.

\subsection{BH mass and mass-to-light ratio}

Figure~3 presents $\chi^2$ as a function of BH mass and \mtl. The
contours are drawn using a two-dimensional smoothing spline (Wahba
1980). As in Wahba (1980), Generalized Cross-Validation determines the
smoothing value; however, the modeled values are relatively smooth and
little smoothing is necessary. We plot only those points near the
$\chi^2$ minimum; we have tried many more models that lie outside the
limits shown in the plot but only highlight the center to show the
contour shape. Models that lie outside these limits are excluded at
much greater than 99\% confidence. Each approximately vertical
sequence represents models with the same ratio of BH mass to galaxy
mass (or \mtl), all of which can use the same orbit library except for
a trivial rescaling of the velocities.

Table~1 presents the properties of the galaxies in this sample. The
columns are galaxy name (col.~1), galaxy type (col.~2), absolute
$B$-band bulge magnitude (col.~3), BH mass (col.~4) and uncertainty
$\sigma_e$ (col.~5), which is defined in Section 4.6, distance in
Mpc (col.~6), mass-to-light ratio and the band (col.~7), central slope
of luminosity density (col.~8), shape of the velocity dispersion
tensor in the central model bin (col.~9), shape of the velocity
dispersion tensor at a quarter of the bulge half-light radius
(col.~10), and half-light radius of the bulge (col.~11).  Bulge
magnitudes come from Kormendy~\& Gebhardt (2001). The half-light radii
come from Faber~\etal\ (1989) and Baggett~\etal\ (2000).

In all but two of the galaxies, there is little covariance between BH
mass and \mtl. The reason is that we are probing those regions where
the BH mass dominates the potential with multiple resolution
elements. Since the stars contribute a small fraction of the total
mass in this region, varying their mass-to-light ratios has little
effect on the enclosed mass. The two cases in which there is some
covariance, NGC~3377 and NGC~5845, have high-resolution kinematic data
from only a single FOS aperture. Thus, they have poorer spatial
sampling inside of the region where the BH dominates the potential.

Figure~4 shows $\chi^2$ as a function of BH mass for NGC~4564. This
plot has been marginalized over \mtl. For this galaxy, we have spectra
at 33 spatial positions. With 13 velocity bins each, we then have 429
kinematic measurements. The velocity profiles have a smoothing width
of about two bins, and thus the number of degrees of freedom is about
210. For NGC~4564, we ran a large number of models in order to inspect
the shape of the $\chi^2$ distribution and its asymptotic shape at
small mass. Near the minimum of the $\chi^2$, there is noise at the
level of $\Delta\chi^2 \approx 0.5$.


\psfig{file=gebhardt.fig4.ps,width=8.5cm,angle=0}
\figcaption[gebhardt.fig4.ps]{$\chi^2$ as a function of BH mass for
NGC~4564 marginalized over the mass-to-light ratio. The inset box is a
magnified view of the region near the minimum. The tick marks along
the ordinate on the inset represent $\Delta\chi^2=1$.
\label{fig4}}



\begin{figure*}[t]
\centerline{\psfig{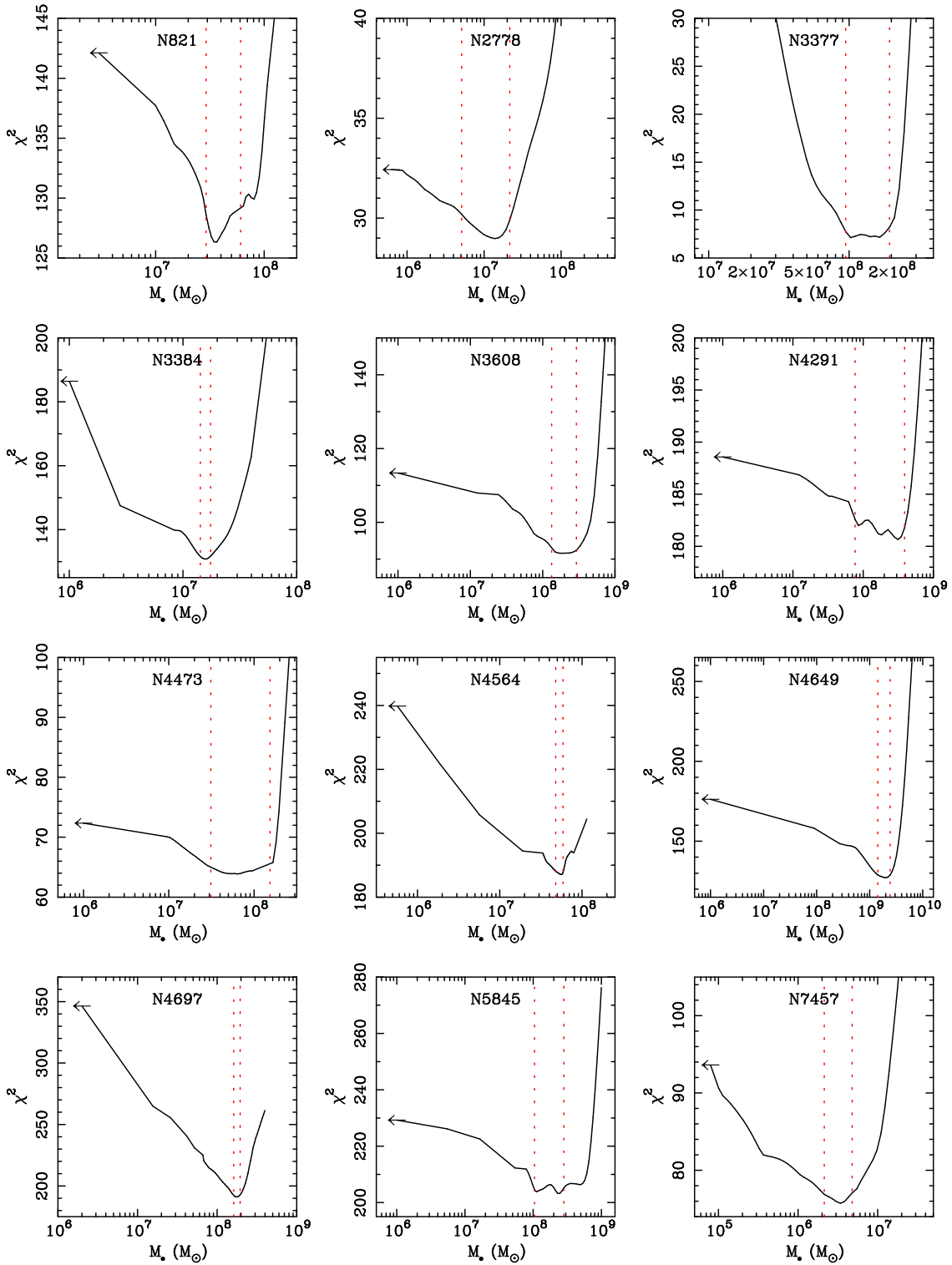}}
\vskip 0pt \figcaption[gebhardt.fig5.ps]{$\chi^2$ as a function of
BH mass for each of the galaxies. We have marginalized over
the mass-to-light ratios.  The vertical dashed lines are the 68\%
confidence bands quoted for the BH mass uncertainties. The
arrow on the leftmost point indicates that this point is actually at
zero BH mass, off the edge of the panel.
\label{fig5}}
\end{figure*}


Figure~5 shows the $\chi^2$ distributions for the whole sample of 12
galaxies. These plots have been marginalized over \mtl\ so $\chi^2$ is
a function of only one variable, $M_{BH}$. In these plots,
$\Delta\chi^2=1$ corresponds to $1\sigma$ uncertainty or 68\%. Thus
the detection of a BH --- or, strictly speaking, of a massive dark
object --- is very significant in most of these galaxies. The least
significant detection is NGC~2778 where the difference in $\chi^2$ is
less than ten between the best-fit BH mass and zero BH mass.

We note the difference in the convention used here for the contour
levels compared to other orbit-based studies. We report uncertainties
that are based on one degree of freedom (i.e., marginalizing over the
other parameters) and are at the 68\% level ($1\sigma$). Other studies
have used different values. Van der Marel~\etal\ (1998) report
$3\sigma$ uncertainties based on two degrees of freedom (BH mass and
mass-to-light ratio) corresponding to $\Delta\chi^2=11.8$. Cretton \&
van den Bosch (1999) report $1\sigma$ uncertainties with two degrees
of freedom corresponding to $\Delta\chi^2=2.3$. Cappellari~\etal\
(2002) and Verolme~\etal\ (2002) report $3\sigma$ uncertainties with
three degrees of freedom (including inclination) corresponding to
$\Delta\chi^2=14.2$. Our intention is to use the BH masses reported
here in galaxy parameter studies, and so we desire a BH mass
uncertainty that has been marginalized over all other
parameters. Furthermore, convention suggests that $1\sigma$
uncertainties are the most useful for parameter studies. Thus, we use
$1\sigma$, 1 degree of freedom uncertainties. We also convert the
uncertainties from the orbit-based studies above to our convention of
$\Delta\chi^2=1.0$. The three galaxies for which we converted the
uncertainties are M32 (Verolme~\etal\ 2002), NGC~4342 (Cretton \& van
den Bosch 1999), and IC1459 (Cappellari~\etal\ 2002). These are
reported in Table~3. In order to do this properly requires sampling
the dynamical models finely enough to see $\Delta\chi^2=1.0$
variations.  Since the above studies were not concerned with the
uncertainties at this level, we must use an approximation. We can use
the $\chi^2$ contours from our sample to approximate the change in BH
mass uncertainty relative to change in $\chi^2$. This is not ideal but
does serve as a first approximation. Thus, $\Delta\chi^2$ changing
from 14.3 to 1.0 implies an average change in the BH mass uncertainty
of a factor of 4 (which we use for M32 and IC1459). Going from
$\Delta\chi^2$ of 2.3 to 1.0 implies an uncertainty change of a factor
of 1.6 (which we use for NGC~4342).

Most of the galaxies in our sample have significant flattenings. Since
it appears that the distribution of intrinsic flattenings peaks at
axis ratio 0.7 (Alam \& Ryden 2002), most of our galaxies should not
be far from edge-on. Except for NGC~4473, all of the models presented
in Figure~5 assume edge-on configuration. We have run a few inclined
models for NGC~3608 and NGC~5845. In both cases, the the BH mass is
within the uncertainty given for the edge-on model.  For the more
face-on configurations, the BH mass increased by 30\% for NGC~3608,
and decreased by 20\% for NGC~5845. Gebhardt~\etal\ (2000a) found that
some inclined models for NGC~3379 had BH masses larger by a factor of
two compared to the edge-on case (but still within the
uncertainty). However, the data used for these studies have limited
spatial kinematic coverage. The two-dimensional kinematic dataset used
for M32 (Verolme~\etal\ 2002) provides the optimal way to study
inclination effects. They find that the more face-on case gives a 30\%
decrease in the BH mass measured from the edge-on model. Since we do
not have adequate angular kinematic data to constrain the inclination,
we rely on the above studies and the few cases that we have run to
determine the inclination effect. On average, it appears that
inclination will cause a 30\% random change in the BH mass. Our BH
mass uncertainties range from 10\% to 70\%, with the most flattened,
nearly edge-on, galaxies tending to have the smallest
uncertainties. For those galaxies where inclination might be a
concern, their uncertainties are larger than 30\%. Thus we do not
include any additional uncertainty that might be caused from using the
incorrect inclination. The uncertainties given in Table~1 include
those as measured from the edge-on models alone (or from the one
inclined case for NGC~4473). Inclined models do, however, affect the
\mtl; as the galaxy approaches a more face-on configuration, it
becomes more intrinsically flat in order to give the same projected
flattening. Since it must maintain a similar projected dispersion,
then the smaller column depth for the more flattened galaxy requires a
higher \mtl, which is what is seen in the models.

The uncertainty in the BH mass determination clearly depends on both
the spatial resolution and the S/N of the data. The spatial resolution
can be parametrized relative to the radius of the sphere of influence,
$GM_{BH}/\sigma^2$. For our galaxies, these radii range from
0.02\arcsec\ in NGC~2778 to 0.75\arcsec\ in NGC~4649. The size of the
central bin used in the modelling is 0.05\arcsec. Thus, for NGC~2778,
the sphere of influence is more than a factor of two below our
resolution limit.  Because of this small radius, we have tried a
variety of different datasets applied to the NGC~2778 models and find
similar results. The zero BH mass model for NGC~2778 is ruled out at
only the 95\% confidence limit --- our least confident detection.  For
all other galaxies, the sphere of influence is larger or equal to our
resolution limit.

The uncertainties in the BH mass come from the shape of the one
dimensional $\chi^2$ contours. It is important to check whether this
estimate of the uncertainties properly reflects the true
uncertainties. We can check this to a limited extent through Monte
Carlo simulations of the kinematics. We note that this study will only
determine the uncertainties within our assumptions; we discuss effects
from relaxing our assumptions in Section 4.9. We use the same Monte
Carlo realizations that were used for the spectra (as described in
Pinkney~\etal). For each realization of the set of LOSVDs we find the
BH mass that provides the minimum $\chi^2$. With 100 realizations, we
then determine the 68\% confidence limits from the Monte Carlo and
compare that to the same limits as determined from the shape of the
$\chi^2$ contour. The uncertainties as measured from both techniques
are in excellent agreement. Assuming that the Monte Carlo simulations
should provide the most accurate uncertainties, we find no reason to
question the uncertainties as measured from the $\chi^2$ shape, due to
their concordance. We have run this experiment only on NGC~3608 but
believe these results to be general.

Gebhardt~\etal\ (2000b) presented preliminary BH masses based on this
analysis. Most of the preliminary masses are the same as those
presented here, except for changes due to the change in assumed
distance ($M_{BH} \propto \hbox{distance}$). The few other differences
arise because we now use a higher resolution grid of models, so the
minimum of $\chi^2$ is located more accurately. The changes in both
the best-fit mass and the uncertainties are generally less than
$0.5\sigma$. The most extreme change is in NGC~5845 since the lower
limit for that galaxy was defined using a very poorly sampled grid.

\subsection{Individual Galaxies}

Pinkney~\etal\ (2002a) provide observational notes for the ten
galaxies in our sample that were observed with STIS. Here we report
any additional details of the dynamical models for our sample of
twelve galaxies. In addition, we include notes for five other galaxies
taken from the literature that have similar models and are used in the
analysis in Sections 4.5 and 4.7.

{\it NGC~821:} There are 312 velocity constraints, coming from 24
spatial positions each with 13 LOSVD bins. Given that 2--3 adjacent
bins are correlated from the velocity profile smoothing, the number of
degrees of freedom is about 100, so the minimum $\chi^2$ of 128
(Table~2) indicates a good fit.

{\it NGC~2778:} There are additional ground-based data from
Fisher~\etal\ (1995) along the major axis.  NGC~2778 is important
because its BH mass is low relative to the $M_{BH}/\sigma$
relation. We have modelled NGC~2778 using three ground-based datasets:
the STIS data plus our ground-based data, the STIS data plus the
Fisher~\etal\ data, and the STIS data plus both ground-based
datasets. All three best-fit BH masses are consistent at the 68\%
confidence level. The data from Fisher~\etal\ have higher S/N and so
the results we present are based on the STIS data plus the
Fisher~\etal\ data.

{\it NGC~3377:} The black hole mass in NGC~3377 was first measured in
Kormendy~\etal\ (1998) using only ground-based data; the mass that we
find here is within their uncertainties. We use ground-based data from
Kormendy~\etal\ (1998) along the major and minor axes. Kormendy~\etal\
present only the first two moments of the velocity distribution. Since
our models require data on the full LOSVD, we convert these moments
into a Gaussian velocity profile. The uncertainties are generated
through a Monte Carlo procedure; we generate 1000 velocity profiles
consistent with the means and uncertainties of the moments. The
uncertainty at each velocity bin is given from the 68\% range about
the mean in the simulations.  From {\it HST}, we have two FOS
observations, which we present in the Appendix. Since we only use the
first two moments to generate the LOSVD and since galaxies can have
significantly non-Gaussion LOSVDs, we have checked whether including
additional moments affects the results. We have included a variety of
H3 and H4 components for the ground-based data, using values that are
consistent with those from other galaxies. We find little difference
in the BH mass as reported in Table~1. The main reason for this is
that the HST data shows a dramatic increase in the central dispersion
and in the rotation relative to the ground-based data. Thus, the BH
mass is determined mainly from the radial change in the kinematics and
not from the higher order moments of the LOSVD.

{\it NGC~3384:} NGC~3384 is one of the two galaxies that show a
smaller velocity dispersion in the {\it HST} data than in the
ground-based data. The reason for this drop is that the STIS
kinematics are coming mainly from a cold edge-on disk. The dynamical
models are free to include as many circular, or nearly circular,
orbits as necessary, and so they easily match the kinematic
profile. NGC~3384 is one of the more significant BH detections.

{\it NGC~4473:} NGC~4473 shows a flattening in the central isophotes
and also a significant decrease in the central dispersion. Both of
these indicate the presence of a central disk (see Pinkney~\etal\
2002a). Central stellar disks are seen in many elliptical galaxies
(Jaffe~\etal\ 1994, Lauer~\etal\ 2002). In order to provide the best
representation for the dynamical models, we include a central
disk. The parameters of the disk are measured from the {\it HST}
images. We use a spheroidal representation for the bulge component and
model the residual with a zero-thickness disk with an exponent of
0.5. The parameters for the exponential disk are
$4.9\times10^7\Lsun/{\rm arcsec}^2$ for the central surface bightness
and 1.0\arcsec\ for the scale length. The best-fit inclination is
72\degr\ which we also assume for the galaxy. The mass of the disk
inside of 1\arcsec\ is 20\% higher than the bulge mass in that
region. Thus, it does have a noticeable effect on the kinematics. The
models have no problem matching the high rotation and low dispersion
of the disk.

{\it NGC~4649:} NGC~4649 is the largest galaxy in our sample and has
the lowest surface brightness. We spent 22 {\it HST} orbits exposing
on this galaxy. The central dispersion, 550~\kms, is the highest ever
observed. Despite the large dispersion and low surface brightness,
both of which strongly affect the S/N, the uncertainty in the BH mass
is only 30\%.

{\it NGC~4697:} There is a gas disk in the center of this galaxy, and
the gas kinematics for this galaxy are measured by Pinkney~\etal\
(2002b). NGC~4697 has the most significant BH detection. The
difference in $\chi^2$ between the zero BH mass model and the best fit
model is 155.

{\it NGC~7457:} There is a central point nucleus in NGC~7457. When
measuring the surface brightness profile, we first subtract a point
source from the center. Thus, for the stellar luminosity density, we
assume that the point source is coming from nonthermal emission and
does not contribute to the stellar density. If the point source is a
nuclear star cluster instead of weak nuclear activity, then we will
bias our BH mass since we would have then ignored some of the stellar
mass. The total light in the point source is substantial, $V \approx
18.1$ mag. This amount of light translates into $1 \times
10^7~\Lsun$. Given the BH mass that we measure of $3.5 \times
10^6~\Msun$, assuming that the point source is stellar is inconsistent
with the STIS kinematics. The other effect that it may have is in the
kinematics since the radius at which the point source is contributing
light may be much smaller than the STIS pixels. Thus, the smaller
radius would imply a smaller BH mass for the same dispersion
measure. We do not have a good way to estimate this effect since it
would depend strongly on the actual size of the assumed point source,
but we can get some feel by comparing results from models using only
ground-based observations. For those data, we measure a BH mass
similar to that when including the STIS data, suggesting that the
point source does not have a dramatic effect on the kinematics.

There are two main observations that suggest that the point source is
nonthermal. First, the STIS kinematics show a significant decrease in
the equivalent widths of the Ca~II triplet lines. The drop is around
40\% suggesting nearly equal contribution from stellar and continuum
sources. This drop is also seen in the ground-based data which had a
spatial FWHM of $\sim 1$\arcsec. Second, the point source is
unresolved at {\it HST} resolution. At 13.2 Mpc, the implied scale for
the source is less than 2 pc. Given the luminosity of the source, this
radial scale implies an extremely dense structure, denser than any
known stellar cluster. These two facts lead us to conclude that the
source is nonthermal and must be excluded from the dynamical
analysis. The most likely explanation is that the point source is a
weak active galactic nucleus. Ravindranath~\etal\ (2001) find nuclear
sources in 40\% of galaxies that they observed with {\it HST}, and
they conclude that most of these are likely weak AGN. Ho~\etal\ (1995)
see no obvious nuclear emission from NGC~7457, and we conclude that
it is most likely a weak BL Lac object. The luminosity density for
NGC~7457 in Figure~1 excludes the central point source.

Below are notes for the other galaxies with orbit superposition models
taken from the literature:

{\it M32:} Verolme~\etal\ (2002) have used both STIS spectroscopy and
high S/N ground-based two-dimensional spectra to provide one of
the best measured BH masses using orbit-based models.

{\it NGC~1023:} The results for NGC~1023 are given by Bower~\etal\ (2001) and
will not be repeated here. The only difference is the assumed distance which
changes both the BH mass ($M_{BH} \propto$ distance) and the mass-to-light
ratio (\mtl\ $\propto$ 1/distance).

{\it NGC~3379:} The data and orbit superposition models are presented
by Gebhardt~\etal\ (2000a). NGC~3379 has only a single FOS pointing
using the 0.21\arcsec\ aperture.

{\it NGC~4342:} Cretton~\& van den Bosch (1999) use seven FOS aperture
pointings and ground-based data along several position angles. The FOS
aperture had 0.26\arcsec\ diameter.

{\it IC 1459:} Cappellari~\etal\ (2002) use both STIS spectra and
extensive two-dimensional ground-based spectral coverage. This galaxy
is very important since it also has a measurement of the BH mass from
gas kinematics (Verdoes-Kleijn~\etal\ 2000). The stellar kinematic
measurement is almost a factor of six higher than the gas
measurement. This discrepancy is far larger than the typical error
from our sample measured from stellar kinematics. However, important
uncertainties attach to gas measurements, such as the orientation of
the innermost gas disk and the assumption that the gas is in perfectly
circular orbits. The large residual here suggests that these
uncertainties perhaps deserve more attention than they have received
to date.


\begin{figure*}[t]
\vskip -80pt\centerline{\psfig{file=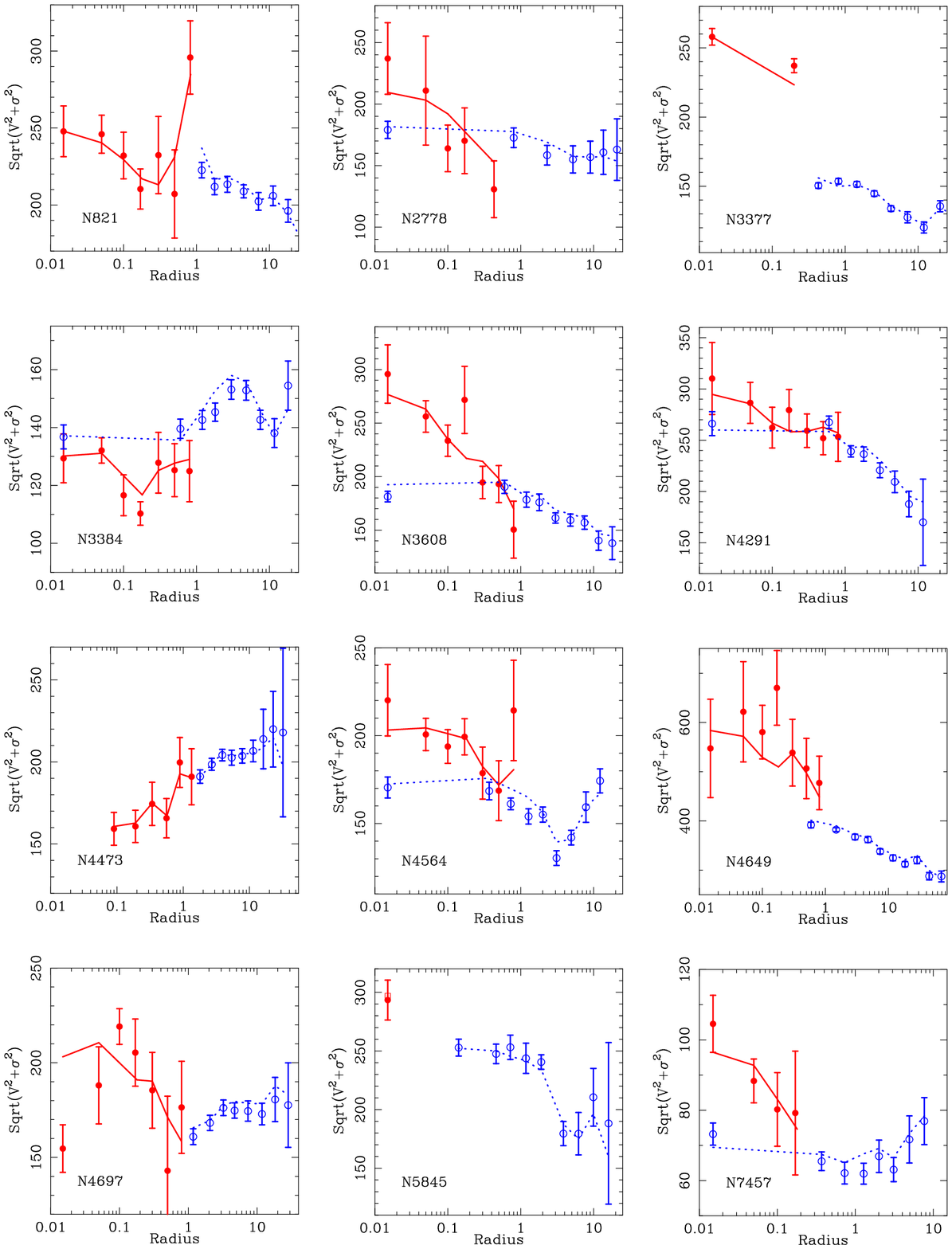,width=17cm,angle=0}}
\vskip -30pt \figcaption[gebhardt.fig6.ps]{The rms line-of-sight
velocity $(V^2 + \sigma^2)^{1/2}$ (in \kms) as a function of radius
(in arcseconds) for each of the galaxies along the major axis. We use
$V$ and $\sigma$ as measured from a Gauss-Hermite fit; since we do not
correct for the higher-order moments, these values approximate the
actual second moment. The red filled circles are the {\it HST} (STIS
or FOS) measurements and the blue open circles are the ground-based
values. The lines are the model results: red solid lines include the
{\it HST} PSF and the blue dotted lines include the ground-based
PSF. For NGC~5845, the model value for the central FOS measurement is
shown as an open square. We provide these plots only for comparison
and note that the models minimize $\chi^2$ using the full LOSVD rather
than its second moment.
\label{fig6}}
\end{figure*}



\begin{figure*}[t]
\centerline{\psfig{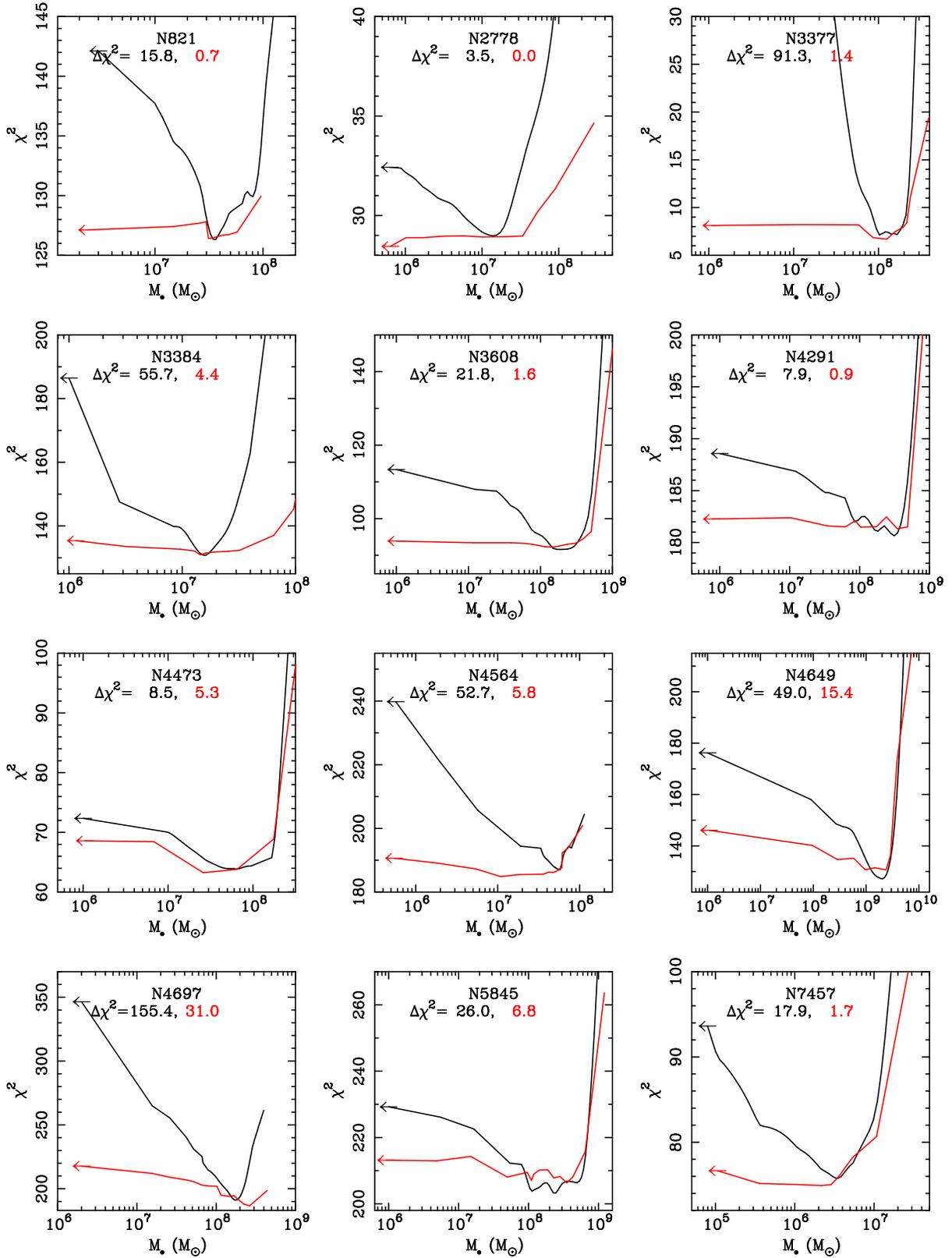}}
\vskip 0pt \figcaption[gebhardt.fig7.ps]{$\chi^2$ as a function of
BH mass using two sets of spectral data for each galaxy. The
solid lines are the same as in Figure~5 where we have included all of
the spectral data (i.e., both {\it HST} and ground-based
observations). The red lines are the results when only including the
ground-based spectral data, excluding the {\it HST} spectra. Below the
galaxy name, we have included the change in $\chi^2$ between the
zero BH model and the best-fit BH model, for both {\it
HST}+ground-based spectroscopy (first entry) and only ground-based
spectral data (second entry). In every case, the significance of the
BH detection is greatly enhanced by including the {\it HST}
spectral data.
\label{fig7}}
\end{figure*}


\subsection{Quality of the Fit}

Figure~6 presents the root-mean-square (rms) line-of-sight velocity as
a function of radius for both the data and best-fit model. [Strictly,
we show $(V^2+\sigma^2)^{1/2}$, where $V$ and $\sigma$ are the mean
velocity and dispersion of the Gaussian that appear in the
Gauss-Hermite expansion of the LOSVD.] We stress that the model uses
the velocity profiles in the fitting and not the second moments
directly. Thus, there are more parameters that control the quality of
the fit than those shown in Figure~6. In addition, some galaxies have
several position angles and we show only one in Figure~6.

For each galaxy, the solid red line and the dashed blue line come from
the same model; the only difference is that they use a different
PSF. For example, the central ground-based measurements for NGC~3377,
NGC~3608, NGC~4564, and NGC~7457 are all significantly different from
the central STIS measurement. This is due to smaller PSF of {\it HST}
which is taken into account in the model. There are two galaxies,
NGC~821 and NGC~4564, which show a spike in the second moment at the
outer STIS radius. This spike is an artifact since we are using only
$V$ and $\sigma$ from the Gauss-Hermite fits, as opposed to including
the higher-order moments, H3 and H4, in the estimate of the second
moment.

In order to judge the quality of the fit, we have to compare $\chi^2$
to the numbers of degrees of freedom (ndof). The ndof is difficult to
measure mainly because there is a smoothing parameter in the
estimation of the velocity profiles; this effect typically decreases
the ndof by a factor of two. An additional difficulty in calculating
the ndof arises since we often include the outer regions of the
velocity profiles where they have no light in the models. Sometimes
these regions extend to velocities that are either outside of those
measured in the velocity profiles or are very uncertain there. In the
regions that are beyond the velocities in which the LOSVDs probe, we
use the uncertainty at the last measured velocity. Since we have no
velocity profile there, we also set the observed LOSVDs in those bins
to zero. Thus, the result is to add zero to the overall $\chi^2$, yet
increase the ndof. The problem can most easily be seen in a galaxy
that has a significant dispersion gradient with radius. For the
modeling, we use a fixed velocity interval and bins for the LOSVD. In
such a galaxy with a large dispersion gradient, the outer edges of the
velocity profile in the center of the galaxy will contain some light,
while those regions at large radii will not. This effect can be
dramatic in some galaxies, causing about half of the velocity bins to
have zero light for the large-radii LOSVD. Thus, there is a further
reduction in the ndof that one needs to apply in order to judge the
quality of the fit. Column~6 in Table~2 reports the total numbers of
fitted parameters for each of the galaxies. Comparing these numbers to
the total minimum $\chi^2$ (Col.~7, Table~2) shows that the $\chi^2$
values are about 2--3 times lower than the ndof, implying reduced
$\chi^2$ values near 0.4. However, this low reduced $\chi^2$ is in
good agreement with the reduction expected from the two effects above.

\subsection{The Need for {\it HST}}

The high-resolution spectral data presented here represent over 100
orbits of {\it HST} time. It is illuminating to determine the
importance of these observations compared to ground-based data.  For
each galaxy, we have re-computed the best-fit models using {\it only}
the ground-based spectra (we still use both ground-based and {\it HST}
photometry). Figure~7 plots the $\chi^2$ as a function of BH
mass for both sets of data (the {\it HST}+ground and ground only). In
every case, inclusion of the {\it HST} data makes a substantial
improvement in the significance of the BH detection (see also
Table~2).

The two galaxies with the strongest BH detection based on the
ground-based data are NGC~4649 and NGC~4697. Of the twelve galaxies in
the sample, these two have the largest angular sphere of influence,
0.75\arcsec\ and 0.4\arcsec\ respectively. The ground-based data come
from MDM where the seeing is typically 1\arcsec. Thus, it is not too
surprising that we can detect the BH in these galaxies without {\it
HST} data. However, when the {\it HST} data are included, in both of
these galaxies the significance is greatly increased.

We can also check whether the BH masses estimated from the two sets of
data are the same. Figure 8 plots this comparison. All of the masses
estimated from the two sets of data are consistent at the 1$\sigma$
level (i.e. all of the error bars in Figure 8 overlap the straight
line). There is no evidence that masses based on ground-based data
alone are systematically high; if anything, the use of ground-based
alone appears to slightly {\it underestimate} the BH mass. A striking
feature of Figure 8 is that even when the 1$\sigma$ uncertainty in the
BH mass from ground-based data includes zero, the best-fit mass from
these data is very similar to the best-fit mass from the full dataset.

Magorrian~\etal\ (1998) presented masses based on ground-based data
and two-integral axisymmetric models. Subsequent analysis shows that
some of the BH masses were overestimated by up to a factor of
three. Merritt~\& Ferrarese (2001) argue that this bias is due to use
of the ground-based data. From the results presented here, it appears
that the problem does not lie in using ground-based data, but more
likely in the model assumptions. For the eight galaxies common to the
present paper and Magorrian~\etal, we find that using our
higher-resolution HST+MDM kinematics has little effect on the BH
masses found by the two-integral models. Therefore the error in the
BH masses of Magorrian~\etal\ is due to their assumption of
isotropy. In particular, the axisymmetric models in this paper exhibit
some radial anisotropy in the velocity-dispersion tensor at mid-range
radii; as Magorrian~\etal\ point out, radial anisotropy will cause the
simpler isotropic models used in that paper to overestimate the
masses. For the 12 galaxies in Magorrian~\etal\ (1998) that have
non-zero BH mass estimates and are also in the Tremaine~\etal\ (2002)
sample, the mean overestimate in $\log M$ is 0.22 dex.


\psfig{file=gebhardt.fig8.ps,width=8.5cm,angle=0}
\figcaption[gebhardt.fig8.ps]{The BH masses measured from all of the
spectral data ({\it HST} and ground-based) compared to those measured
from using the ground-based spectral data only. The confidence bands
are 68\% uncertainties. All measurements based on ground-based data
alone are consistent with measurements based on {\it HST}+ground-based
data. The masses from the ground-based spectral data may tend to
slightly underestimate those obtained including the {\it HST} spectral
data.
\label{fig8}}



\begin{figure*}[t]
\centerline{\psfig{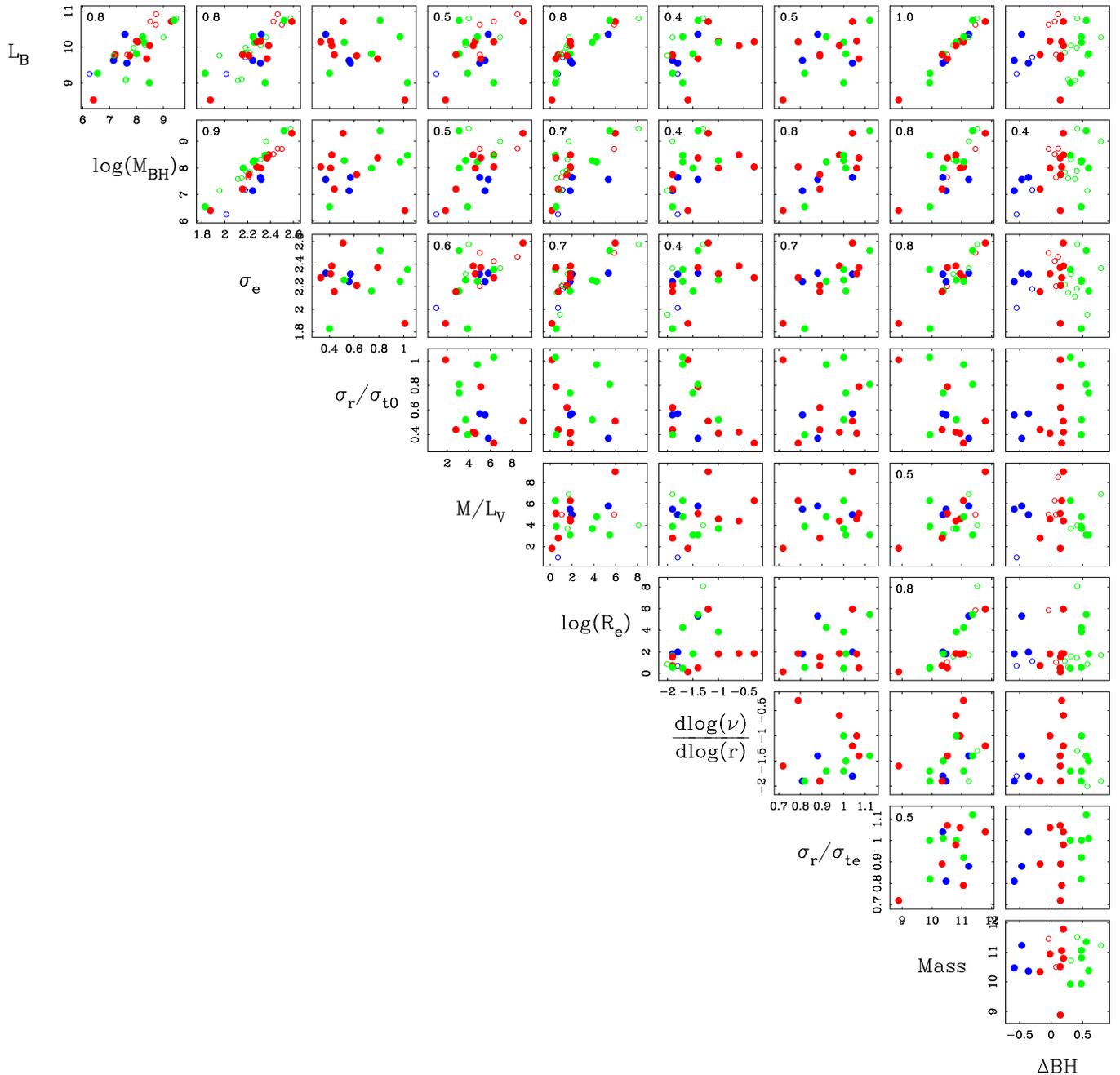}}
\vskip 0pt \figcaption[gebhardt.fig9.ps]{Ten galaxy properties plotted
against each other, for the galaxies that have measured BH masses. For
this plot, we include up to 31 galaxies; the BH masses are the same as
in Tremaine~\etal\ (2002). The colors represent the offset from the
$M_{BH}/\sigma$ correlation as defined in Tremaine~\etal, with green
having the largest positive BH offset, blue having the largest
negative offset, and red having small offset. The number written in
the upper left corner of the plot is the Pearson's R correlation
coefficient. If the probability from the correlation is below 10\%, we
do not report R. Filled symbols are the 17 galaxies that have orbit
based models; the rest are plotted with open symbols.
\label{fig9}}
\end{figure*}


We have also investigated whether reliable BH masses can be obtained
from {\it HST} spectral data alone, excluding the ground-based
spectra. We ran the models on the one galaxy that should have produced
the strongest BH detection based on {\it HST} alone. NGC~3608 shows a
dispersion increase by a factor of two just in the {\it HST} data,
from 0\arcsec--1\arcsec. For these data, we find {\it no} significant
detection for a BH, suggesting that the ground-based data are
necessary to measure one. The reason is that the stellar \mtl\ is
unconstrained by the {\it HST} data alone. The \mtl\ implied for
NGC~3608 from the {\it HST} data is about a factor of two higher than
that found when using all of the data together. This increase in the
\mtl\ causes the significance of the BH detection to disappear in the
{\it HST} data alone.


\begin{figure*}[t]
\centerline{\psfig{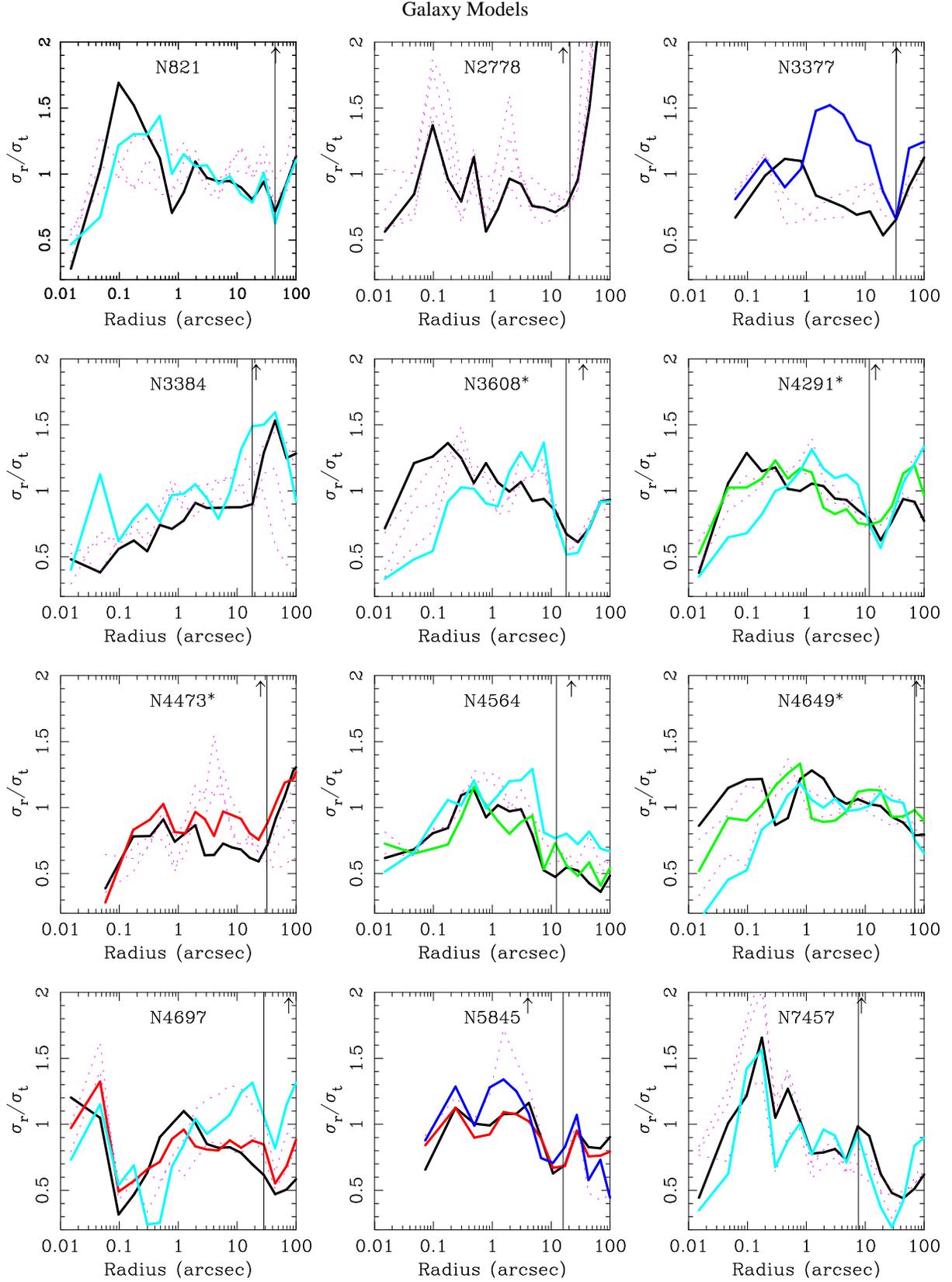}}
\vskip 0pt \figcaption[gebhardt.fig10.ps]{The shape of the velocity
dispersion tensor for all twelve galaxies, plotted as a function of
radius. The solid lines are at those position angles for which we have
kinematic data. The dotted lines are those for which we do not have
data but result from the maximum entropy solution for the best-fit
model. The colors represent different position angles. The black line
is along the major axis and the light blue line is near the
pole. Other colors are intermediate axes. The vertical solid line is
the radial extent of the ground-based data, and the arrow is at the
half-light radius. An asterisk denotes whether the galaxy is a core
galaxy.
\label{fig10}}
\end{figure*}


\subsection{Black Hole Correlations with Galaxy Properties}

We are now in a position to compare BH masses with other host galaxy
properties, to look for underlying relationships that may inform us
about the formation process of the BH. In Figure~9, we plot ten galaxy
properties, including the BH mass, against one another. For example,
in the first plot on the left on the top row, we plot the BH mass
along the abscissa and the bulge luminosity along the ordinate. In
addition to the galaxies with BH masses measured in this paper, we
have added other galaxies with reliable mass estimates, for a total
sample of 31 galaxies.  Tremaine~\etal\ (2002) report some of the
properties of the galaxies not included here. In Figure 9, the number
of galaxies in each panel changes depending on whether that particular
value exists for all 31 galaxies. However, we do differentiate between
those galaxies studied with orbit-based models (filled symbols) and
those with other models (open symbols).

The galaxy properties that we report are bulge luminosity, BH mass,
effective dispersion (discussed below), radial to tangential
dispersion at the galaxy center (discussed below), mass-to-light ratio
in the $V$ band, bulge half-light radius $R_e$, central luminosity
density slope ($d\log\nu/d\log r$, where $\nu$ is the luminosity
density), radial to tangential dispersion on the major axis at a
quarter of the half-light radius, bulge stellar mass, and BH mass
offset from the $M_{BH}/\sigma$ correlation. The bulge luminosities
are taken from Kormendy~\& Gebhardt (2001) and the calculations are
given by Kormendy~\etal\ (2002). The bulge half-light radii come from
Faber~\etal\ (1989) and Baggett~\etal\ (2000). The BH mass offset is
calculated using the relation in Tremaine~\etal\ (2002). For the
mass-to-light ratios, we use only those galaxies that have a measured
value in the $V$ band.  To measure the total mass, we use the bulge
total $B$-band light, convert to $V$ using $B-V$ from RC3
(de~Vaucouleurs~\etal\ 1991), and multiply by the $V$-band
mass-to-light ratio.

There are five galaxies that have axisymmetric orbit-based models from
previous studies. We include the internal velocity structure of these
five galaxies in Figure~9 and subsequent analysis. Table~3 reports
their internal velocity moments and the reference.

Besides the obvious and expected correlation between total mass and
total light, the most significant correlation is the $M_{BH}/\sigma$
relation reported by Gebhardt~\etal\ (2000b) and Ferrarese~\& Merritt
(2000). Tremaine~\etal\ (2002) discuss the differences in the measured
slope of this relation. Most other significant correlations between
various galaxy properties and the BH mass can be regarded as a result
of well-known correlations of other galaxy properties with
dispersion. There is a significant correlation between the shape of
the dispersion tensor at $R_e/4$ and the BH mass. Larger BHs tend to
live in galaxies that have more radial motion. This trend may be a
clue to the formation process of the BH but, also, could represent a
secondary correlation between galaxy anisotropy and galaxy
dispersion. There is a suggestion that larger BH mass offsets occur in
galaxies that have less radial energy near the center, possibly
signifying additional evolutionary effects. However, we need more data
to decide on the significance, since it is only at the 20\% level in
the anisotropy. A full treatment of the correlations should include a
proper principal component analysis (PCA); however, given the
uncertainties and small sample, we are not in a position to explore
PCA, especially since the $M_{BH}/\sigma$ relation provides such small
scatter already.

\subsection{Effective Dispersion}

We use the effective dispersion, $\sigma_e$, as a representation of
the galaxy velocity dispersion; $\sigma_e$ is the second moment of the
velocity profile integrated from $-R_e$ to $+R_e$ along the major axis
with a slit width of 1\arcsec. The idea is to represent the galaxy by
one dispersion estimate. There are many ways to do this; for example,
J\o rgensen~\etal\ (1996), Faber~\etal\ (1989), and the Sloan Survey
(Bernardi~\etal\ 2002) use the second moment inside a circular
aperture of radius $R_e$/8. Tremaine~\etal\ (2002) discuss the effect
that the BH can have on either the effective dispersion or the
dispersion inside $R_e$/8. In most cases, the BH has little effect,
$<$3\%, but in some galaxies the effect can be as large as 30\%.

The effective dispersions are given in Table~1. In all cases the S/N
is very high, over 100. The corresponding statistical uncertainty in
$\sigma_e$ ranges from 1--3\%. However, at this level, systematic
uncertainties dominate, particularly continuum estimation and template
mismatch. We have investigated both of these effects by varying the
continuum level and using different templates.  We find that at any
S/N, using an appropriate range of systematic variables, the overall
uncertainty is no better than 5\%. Therefore, we adopt 5\% accuracy
for the effective dispersion measurements. We discuss below how this
choice affects the main results.

\subsection{Velocity Dispersion Tensor}

We show the shape of the velocity dispersion tensor in Figure~10. We
define the tangential dispersion as $\sigma_t =
[(\sigma_\theta^2+\sigma_\phi^2)/2]^{1/2}$, so that for an isotropic
distribution the radial and tangential dispersions are equal. Note
that $\sigma_\phi$ includes both random and ordered motion (i.e., it
represents the second moment of the azimuthal velocity relative to the
systemic velocity, not relative to the mean rotation speed). The most
obvious trend in Fig~10 is that the tangential motion tends to become
more important towards the center (discussed below) in all galaxies
except for NGC~4697. 


\vskip -180pt \hskip -100pt\psfig{file=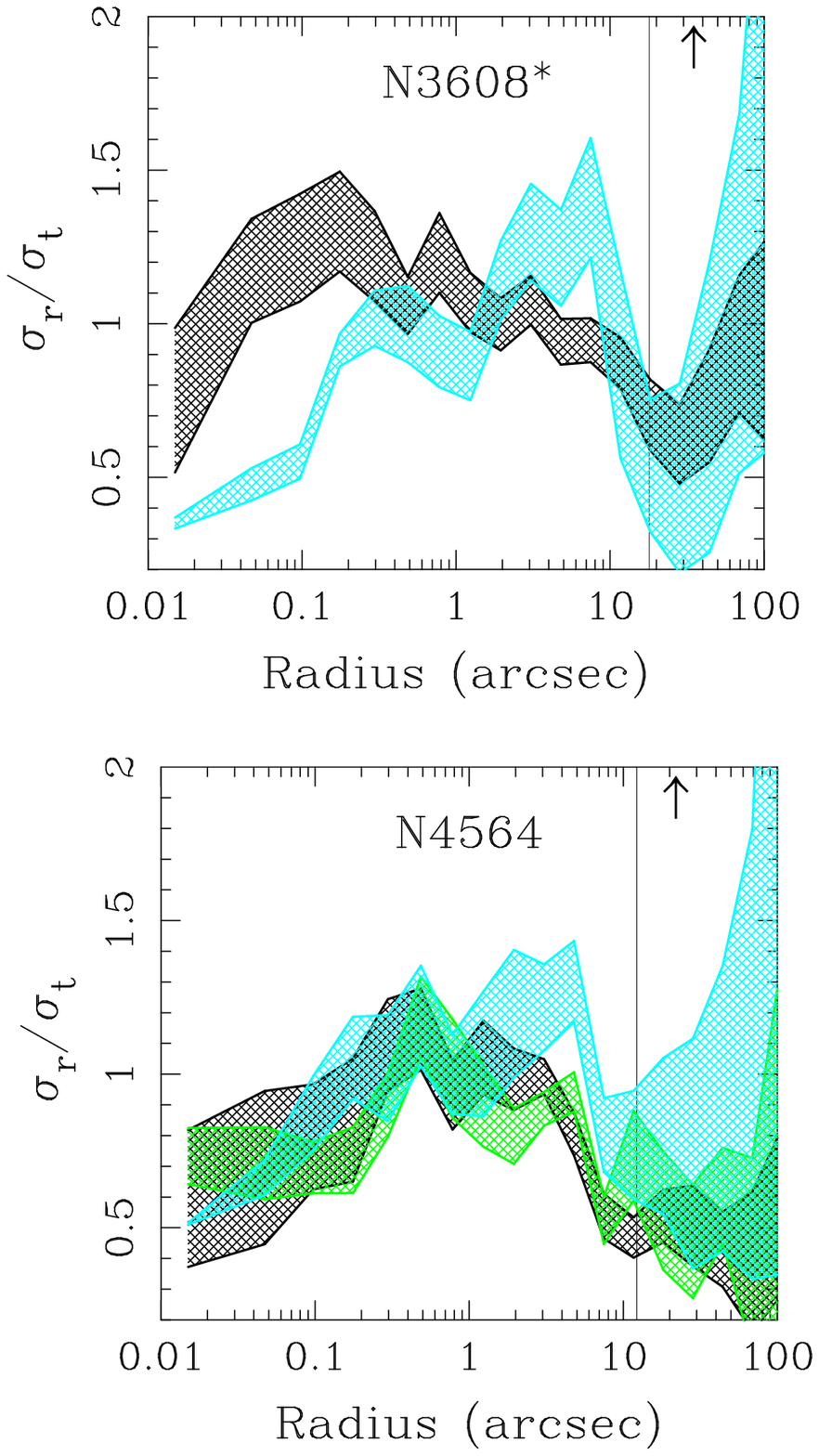,width=15cm,angle=0}
\vskip -100pt
\figcaption{Same as Figure~10, except here we include the
68\% confidence bands based in Monte Carlo simulations for two of the
galaxies. The colors refer to different position angles, as in
Fig.~10. The colored hatched regions refer to their 68\% confidence
bands. The average uncertainty is around 0.15 in $\sigma_r/\sigma_t$.
Beyond the region where we have no kinematic data (the vertical solid
line), the uncertainties increase dramatically.}
\vskip 10pt


We use two methods to measure the uncertainties on these quantities.
We have run Monte Carlo simulations for two galaxies, and for the
remaining we use the simple alternative of using the smoothness in
both the radial and angular profiles to estimate the
uncertainties. For the Monte Carlo simulations we use the same
realizations that are used to generate the LOSVD uncertainties
(Pinkney~\etal) and to generate the BH mass uncertainties. For each
set of LOSVD realizations, we find the best-fit model and examine the
orbital structure. We have run these simulations on two galaxies: the
core galaxy NGC~3608 and the power-law galaxy NGC~4564. Figure~11
plots the results. For the 100 realizations we estimate the 68\%
confidence bands by choosing the 16\% and 84\% values in the sorted
internal moments at each radii. Figure~11 shows that the drop in the
radial motion near the center is statistically significant for both
galaxies. At the radii outside of our last measured kinematic
measurements, the uncertainties become very large, as expected.  We
note that the radial profiles are not very smooth. The level of
non-smoothness is consistent with the measured uncertainties.  This
noise is likely a result of using the same orbit library with each new
LOSVD realization. Thus, the remaining noise is due to using a limited
number of stellar orbits. Ideally, we should include a random sampling
of the photometry, and hence the stellar potential in the Monte Carlo
simulations. By doing so, we would average over noise from a
particular set of orbits. However, this is computationally prohibitive
and we rely on the present simulations to provide the uncertainties.

For the other galaxies, we use deviations from smoothness as an
estimate of the internal orbital structure uncertainties. An
expectation is that the radial and angular gradient of the internal
moments may be smooth, albeit details due to recent merger and
accretion history may cause small scale variations. Thus, deviations
from a smooth profile may be indicative of the measurement
uncertainty. The three galaxies with the smallest number of kinematic
measurements---NGC~2778, NGC~3377 and NGC~7457---show the largest
radial and angular variations, suggesting that these are due to
increased uncertainties from not having as much kinematic constraints.
By inspection of variations seen Fig.~10 and Fig.~11, we estimate that
the uncertainties on ratio of the internal moments is around 0.1 to
0.2 for most galaxies. This uncertainty is also consistent with the
angular variations. The models are free to have very different
dispersion ratios at different position angles and radii. The fact
that the ratios are similar at different angles suggests that the
measurements are robust for these models.  The core galaxies (denoted
with an asterisk) show a larger decrease in the ratio towards the
center which we discuss next.

Although the models produce the internal moments everywhere in the
galaxies (i.e., Fig.~10), Figure~12 shows them only along the major
axis and at two radii: the central bin in the models and an average of
the three bins nearest $R_e/4$. The BH dominates the potential in the
central bin.

Figure~12 shows that galaxies with shallow central density profiles
have orbits with strong tangential bias near their centers. At larger
radii, the orbits tend to be isotropic or slightly radial.  There is a
concern that this change in dispersion ratio may simply reflect our
assumption that the mass-to-light ratio is independent of radius. If a
dark halo were present, so that the mass-to-light ratio increased
outwards, a galaxy with isotropic orbits will appear to become
tangentially anisotropic at large radii. However, it is unlikely that
the dark halo makes a significant contribution to the potential within
$R_e/4$, and in any case the sign of this trend (increasing tangential
anisotropy with radius) is opposite to the one we observe.  Figure~12
only includes results from radii that are small enough to be
unaffected by the presence of a dark halo. In any event, the most
likely bias is that, by not including a dark halo, we will
overestimate the amount of tangential anisotropy at large radii. At
small radii, the dark halo assumption will have no effect. Therefore,
we are confident of the gradient seen in Figure~12.


\vskip 5pt \psfig{file=gebhardt.fig12.ps,width=8.5cm,angle=0}
\figcaption[gebhardt.fig12.ps]{{\it Top}: ratio of radial to
tangential rms velocity, plotted as a function of central luminosity
density slope. Each galaxy appears twice in the plot: the open symbols
represent the ratio at $R_e$/4, the filled symbols are the ratio in
the central bin. An isotropic dispersion tensor corresponds to
$\sigma_{\rm r}/\sigma_{\rm t}=1$. In addition to the 12 galaxies in
this paper, we include an additional five presented in Table~3. The
division between core and power-law galaxies is around
dlog\,$\nu$/dlog\,r = --1.3, with core galaxies having flatter
profiles. {\it Bottom}: the change in shape of the velocity dispersion
tensor (ratio of radial to tangential motion) from the center to
$R_e$/4 as a function of central luminosity density slope.}



\begin{figure*}[t]
\centerline{\psfig{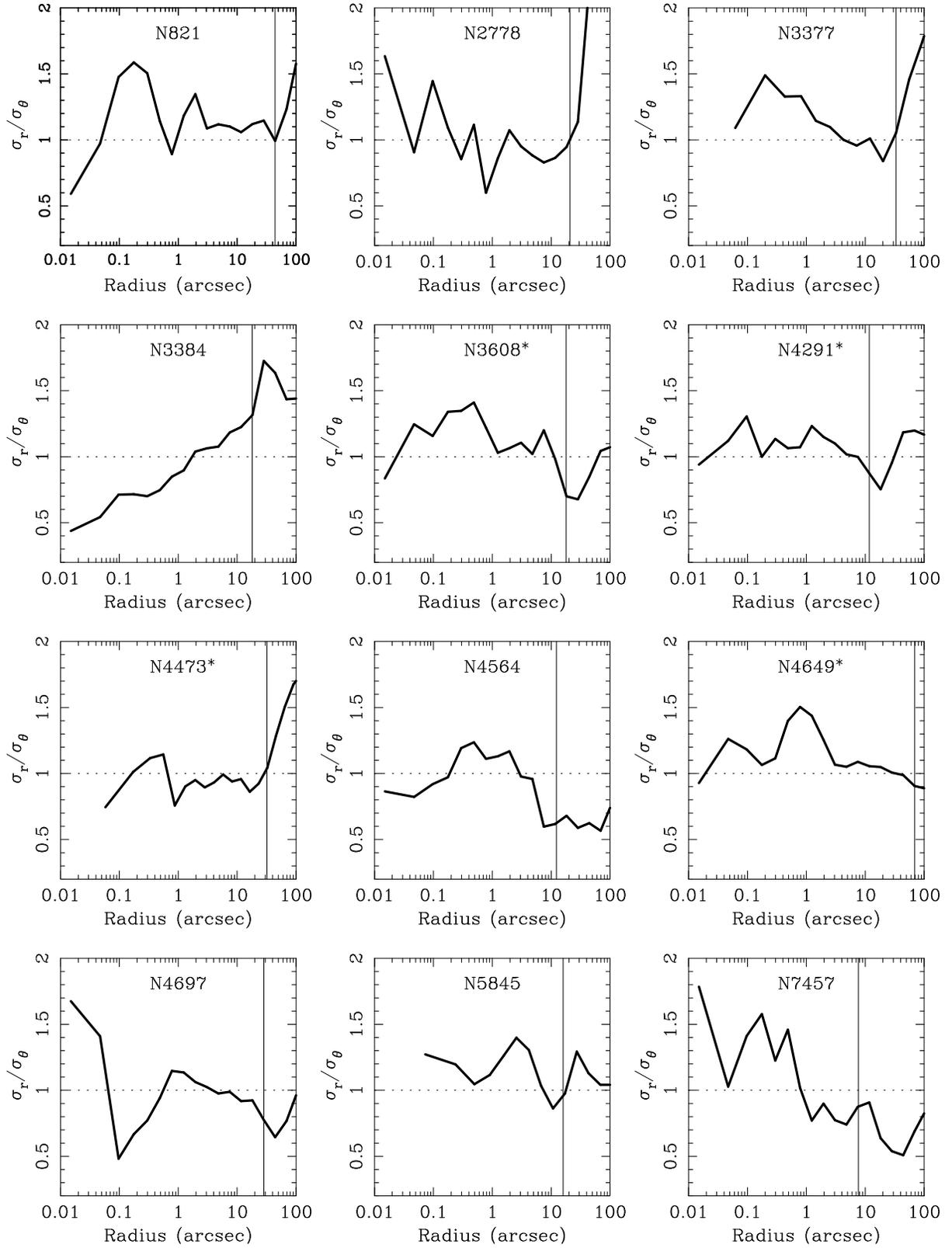}}
\vskip 0pt \figcaption[gebhardt.fig13.ps]{The ratio of the radial to
the $\theta$ velocity dispersion for all twelve galaxies plotted as a
function of radius along the equator for the best-fit model (solid
line). The dotted line is at $\sigma_r/\sigma_\theta = 1$, which is
the result along the major axis for a two-integral model. The galaxies
show a range of profiles with no obvious radial trends. The vertical
solid line is the radial extent of the ground-based data; results
beyond that radii are not meaningful. An asterisk denotes whether the
galaxy is a core galaxy.
\label{fig9}}
\end{figure*}


The bottom panel in Figure~12 plots the change in the
radial-to-tangential motion between the center and $R_e/4$ as a
function of central slope. This plot reiterates that the gradient in
shape of the dispersion tensor is larger for core galaxies than for
power-law galaxies.

The shape of the velocity dispersion tensor depends on the galaxy
formation process. Tangentially biased orbits at small radii can occur
through the destruction or ejection of stars on high-eccentricity
orbits that pass near the BH. There are now 17 galaxies for which we
have measurements of orbital anisotropies; most come from the
dynamical models presented here, but similar results are found using
other orbit superposition studies (M32: van der Marel~\etal\ 1998;
Verolme~\etal\ 2002; NGC~4342: Cretton \& van den Bosch 1999; and
IC~1459: Cappellari~\etal\ 2002). Spherical theoretical models predict
a range of both central cusps and anisotropies. Three models that have
been studied are adiabatic BH growth (Quinlan~\etal\ 1995), fall-in of
a single BH (Nakano~\& Makino 1999), and BH binary models (Quinlan~\&
Hernquist 1997). The adiabatic models grow the black hole by slow
accretion of material (gas or stars). The BH fall-in models start with
a galaxy without a black hole and then a BH is placed a large radii
where is falls in due to dynamical friction. The BH binary model
assumes the galaxy has an existing black hole at the center and then a
second black hole falls in due to dynamical friction, and they
subsequently form a binary BH. The models predict a different value of
the central anisotropy: single BH infall models have
$1>\sigma_r/\sigma_t > 0.87$, adiabatic models have $\sigma_r/\sigma_t
\approx 0.87$, and BH binary models have $\sigma_r/\sigma_t \approx
0.7$. The increased tangential anisotropy for the binary models is due
to the orbital motion of the binary in the galaxy core, which causes
it to affect more stars on radial orbits.

The models that agree best with both observed anisotropies and density
slopes depend on the type of galaxy, whether it is a core or
power-law. Examining the solid points in Figure 12, one notices that
those galaxies with shallow central densities have the largest
tangential motion. The central $\sigma_r/\sigma_t$ for core galaxies
is around 0.4 (highly tangentially biased), while that for the
power-law galaxies range from 0.45 to 1.05 with an average of
0.8. Thus, it appears that the core galaxies are more consistent with
the BH binary models, and the power-law galaxies are more consistent
with adiabatic growth. These conclusions are similar to those of
Faber~\etal\ (1997) and Ravindranath~\etal\ (2002).  Unfortunately,
these comparison models are limited by their simplistic initial
conditions (i.e., isotropic velocity dispersion tensor, spherical
potential). Fortunately, the large number of researchers working in
this area now (e.g., Milosavljevic~\& Merritt 2001;
Holley-Bockelmann~\etal\ 2002; Sellwood 2001) will provide improved
theoretical comparisons.

\subsection{The Need for Three Integrals}

It is important to know whether the distribution function of these
galaxies depends on three integrals of motion or only on the two
classical integrals (energy and $z$-component of angular momentum). In
two-integral models the dispersion in the $R$ and $z$ directions must
be equal (where $R$, $\phi$, $z$ are the usual cylindrical
coordinates). Figure~13 plots this ratio for the 12 galaxies.  There
are no obvious radial trends. For seven galaxies, at small radii the
radial dispersion is higher than in the $\theta$ direction.  For two,
the $\theta$ motion significantly dominates there. At other radii, the
results show a variety of trends with some galaxies having rising
ratios while others having falling ones. The {\it average} ratio along
the major axis is close to unity (i.e., a two-integral model), but the
radial run demonstrates that the best fit model is inconsistent with
having only two integrals of the motion. This result is similar to
that found in other orbit-based models (Verolme~\etal\ 2002;
Cappellari~\etal\ 2002).

We discussed that most of the galaxies (10 of the 12) have substantial
tangential motion near their centers, but not whether this is due to
the $\theta$ or $\phi$ motion. By comparison of Fig~13 to Fig~10, we
notice that at most of the radii, the curves are similar which implies
that the $\theta$ and $\phi$ dispersion are similar. However,
Figure~13 shows a significant increase in the contribution from the
$\theta$ direction near their centers for only two of the galaxies,
whereas from Fig~10, we see that most galaxies show a dramatic
increase in the tangential motion near their centers. Thus, the
dominant component in nearly all of the galaxies near the center is in
the $\phi$ direction. At the center, the $\phi$ dispersion generally
has similar contributions from random and ordered motion. The
theoretical models discussed in \S4.7 do not provide the difference
between the $\theta$ and $\phi$ dispersions, but these could
potentially be important constraints.

\subsection{Possible Concerns}

Our models are limited to axisymmetry. Triaxial and non-symmetric
structures may be common attributes of galaxies. Of the 12 galaxies in
this sample, at least four---NGC~3377, NGC~3608, NGC~4473, and
NGC~7457---show signs of non-axisymmetric structure in the
kinematics. An incorrect assumption of axisymmetry could bias our
results. For example, if a bar is observed down its long axis, the
radial streaming motions along the bar may increase the projected
velocity dispersion. This measured increase may mimic that expected
with a central mass concentration (Gerhard 1988). We have not
investigated the effects of triaxiality on BH mass determinations in
detail, but believe that these effects average to zero when the system
is viewed from many different orientations; thus, triaxiality may
contribute to the scatter in our mass determinations but should not
produce a systematic bias. Furthermore, the scatter due to not
considering triaxiality may be a function of galaxy size, since large
core galaxies possibly are more triaxial than the smaller power-law
galaxies. Clearly, triaxial models should be used to quantify these
effects. Any bias caused by using an inappropriate stellar
distribution could be more dramatic if we only had data at larger
radii where the stellar potential needs to be included. However, as
seen in Figure~8, the ground-based data alone do a fairly good job at
measuring the BH mass compared to when including the {\it HST}. Thus,
at least at the level of our uncertainties, the BH mass is unaffected
by using the large radial data, suggesting that if the galaxies are
not axisymmetric then either the non-axisymmetry is unimportant for
the modeling or it is constant with radius. The best way to test
biases with the axisymmetry assumption is to either model the same
galaxies with triaxial codes or to run the axisymmetric code on an
analytic triaxial galaxy.

We have assumed that the surfaces of constant luminosity density in
all of our galaxies are similar spheroids (with the exception of
NGC~4473, where we add a disk component). This assumption is
consistent with the observation that the ellipticity of the
surface-brightness distribution is similar at all radii, but other
density distributions are also consistent with this observation. The
question is whether the assumption of spheroidal equidensity surfaces
could bias our BH mass determinations. Some guidance comes from the
analysis of Magorrian~\& Ballantyne (2001), who study the influence of
embedded stellar disks. In this case, they find that face-on disks in
round galaxies in projection may bias a spherical model toward having
radial anisotropy. This effect is primarily seen at large radii and is
unlikely to bias the BH mass since we are measuring the kinematics so
close to the BH. But, this effect may be important for the orbital
structure that we measure in these galaxies. Again, at small radii,
the influence of a disk is likely to be small since we do not see
strong signatures of one (except in NGC~4473), however it would be
difficult to measure a disk at larger radii. Thus, there is a concern
that we may be biased by this effect in some of the galaxies at larger
radii. Fortunately, only four of our galaxies are rounder than E3, so
this is unlikely to alter the overall conclusions. There are many
other possibilities other than embedded stellar disks that can lead to
non-uniqueness in the deprojection.  The best way to understand their
effects is to run models with different deprojections. For four of the
galaxies, we have tried a variety of inclinations and find
insignificant changes to both the BH mass and orbital structure.

We have not included a dark halo in this analysis. It appears that in
most elliptical galaxies the dark halo becomes important at about the
half-light radius (Kronawitter~\etal\ 2000; Rix~\etal\ 1997). Even
though we have data and model results at these large radii, and they
are plotted in Figure~10, we do not use the model results from these
radii because they may be seriously comprised by the exclusion of a
dark halo. At radii less than $R_e/4$ the stars and central BH
dominate the potential. For Figure~12, we choose $R_e$/4; at
this radius the stellar potential dominates. The BH mass is
determined almost exclusively by the small-radii data; thus, we are
confident that exclusion of a dark halo is unimportant for the BH 
mass estimate.  Gebhardt~\etal\ (2000a) include various dark-halo
profiles and find little difference in the results inside of
$R_e$/2. The next step in the data analysis is to run models in order
to measure both the BH mass and dark halo properties.

We have assumed that the mass-to-light ratio is constant with radius.
As we discuss above, the exclusion of a dark halo is unlikely to
affect either the black hole mass or the orbital structure, however,
variation at small radii can have an effect. For example, a dramatic
increase in the stellar mass-to-light ratio in the central regions can
decrease the measured BH mass if not accounted for. We have not done a
detailed spectral analysis to determine the stellar makeup but we can
use the color gradients to provide some constraints. For the 12
galaxies, the {\it largest} mass-to-light variation from 10\arcsec\ to
the center is $V-I=0.1$; the average is around 0.04. Models of Worthey
(1994) suggest that a $V-I=0.1$ imply an mass-to-light change of about
20\%. We do not include that small variation here but note that
Gebhardt~\etal\ (2000a) use an even larger variation and find no
change in the measured BH mass. Cappellari~\etal\ (2002) find a
similar result for IC1459. Thus, we conclude that inclusion of a small
mass-to-light variation at small radius will have insignificant effect
on the BH mass.

\section{Conclusions}

The twelve galaxies in this paper all have significant BH detections,
with a typical statistical significance in the masses of around
30\%. The average significance of detection is well above 99\% and the
least significant detection (NGC~2778) has about 90\%
confidence. Thus, for this sample, every galaxy has a BH. In fact,
only one nearby galaxy with high-resolution spectral data lacks any
significant BH detection: the pure disk galaxy M33 (Gebhardt~\etal\
2001). The most obvious difference between M33 and the galaxies with
significant BH detection is that the latter have a bulge component.

For a few of these galaxies, ground-based spectra alone yield
reasonably precise BH masses. The masses based on ground-based data
alone are generally remarkably close to the masses based on
ground-based and {\it HST} data; there is no evidence that masses
based on ground-based data alone are systematically high. The most
important aspect of using ground-based data is assure that the models
are fit using full generality (i.e., without assumptions about the
orbital structure).

The most significant correlation with BH mass is with the velocity
dispersion. The present intrinsic scatter is around 0.23 dex in BH
mass (Tremaine~\etal\ 2002). It will be extremely illuminating to
include more galaxies at both extremes, the low mass and high mass
ends. The next most significant correlation is with the
radial-to-tangential velocity dispersion at $R_e$/4. We do not know
whether this is simply a secondary correlation due to that with the
velocity dispersion, or if it represents an evolutionary pattern due
to the growth of the BH. Detailed theoretical and N-body models are
required to understand this. The BH mass also significantly correlates
with both galaxy bulge luminosity and bulge mass, but neither of these
is as strong as with dispersion.

The uncertainties in the BH masses reported here are only
statistical. We have not attempted to include uncertainties from the
assumptions in our models or systematic errors in our analysis
outlined in \S4.9. We believe that the increase in the uncertainties
is likely to be small, but additional tests are required in order to
substantiate this. We can use the $M_{BH}/\sigma$ correlation as an
approximate constraint on the uncertainties. If there is an underlying
physical mechanism that causes a {\it perfect} correlation between
$M_{BH}$ and $\sigma$, then any scatter seen in the correlation must
be measurement error. Since the current scatter is comparable to the
measurement error, we probably have a reasonable estimate of our
uncertainties; any additional uncertainties caused by our assumptions
should be smaller than 0.23 dex in BH mass. However, this argument
applies only to random errors. If, for example, galaxies deviate from
our assumptions systematically, then the $M_{BH}/\sigma$ correlation
may still have small scatter but incorrect BH masses. The only way to
test this is to include a larger sample with general dynamical models
that cover a wide variety of input configurations.

The orbit-based models provide a look into the internal orbital
structure of an axisymmetric system. Based on the small sample of
galaxies shown here and the limited theoretical comparisons, we are
already able to place some constraints on the possible evolutionary
history of the galaxy. The results in this paper suggest that core
galaxies have tangentially biased orbits near their centers, while
power-law galaxies show a range of tangential relative to radial
motion. As suggested by Faber~\etal\ (1997) and Ravindranath~\etal\
(2002), it appears that the core galaxies are consistent with the
BH/binary models, and the power-law galaxies are more consistent with
adiabatic growth. This conclusion comes from analysis of the stellar
surface brightness profiles, and now a similar conclusion comes from
the stellar kinematics. Significant improvement in our understanding
of the orbital structure will come from datasets with two-dimensional
kinematics. De~Zeeuw~\etal\ (2002) and Bacon~\etal\ (2002) present
examples of datasets that can be exploited for this analysis. However,
in order to make progress in this area we must understand possible
systematic biases that can arise from various assumptions (e.g., dark
halo, different deprojections, lack of axial symmetry, etc.)

\acknowledgements

K.G. is grateful for many interesting discussions with Tim de~Zeeuw,
Ellen Verolme, and Christos Siopis. We thank the referee, Tim
de~Zeeuw, for excellent comments which improved the manuscript. This
work was supported by {\it HST} grants to the Nukers (GO--02600,
GO--6099, and GO--7388), and by NASA grant
G5--8232. A.V.F. acknowledges NASA grant NAG5--3556, and the
Guggenheim Foundation for a Fellowship.

\appendix

\section{FOS and Ground-Based Data for NGC~3377 and NGC~5845}

The ground-based velocities and dispersions for NGC~3377 come from
Kormendy~\etal\ (1998) and will not be repeated here. Since our models
use the LOSVDs, we convert from the first two moments to the velocity
profile using Monte Carlo simulations. Each velocity profile
realization is a Gaussian with the mean chosen from a random draw from
the measured mean using its uncertainty, and the sigma chosen from a
random draw from the measured dispersion using its uncertainty. This
procedure does not take into account the H3 and H4 components that are
likely to be non-zero. However, the model results depend very little
on the higher order moments, since it is mainly the radial run of the
first two moments that determine the BH mass. The height of the LOSVD
at a given velocity is then the mean of the simulations and the
uncertainty is given by the 68\% confidence bands of the simulations.

For NGC~5845, the data were taken with the MDM telescope. The
observational setup and reductions are similar to those outlined in
Pinkney~\etal\ (2002a). We used the Ca~II triplet region around
8500~\AA. The plate scale is 0.59\arcsec\ per pixel. The wavelength
scale is 1.44~\AA\ per pixel, with an instrumental resolution of
0.75~\AA\ or 26~\kms. We observed along three position angles for
NGC~5845: 0\degr, 22\degr, and 90\degr\ (defined from the major axis
to the minor axis), with total exposure times of 3 hours for each
position angle (9 hours in total). In Table~4, we report the first
four velocity moments of the LOSVD for the three position angles.
They are plotted in Figure~A14.

For the FOS data, the reduction procedure is similar to that in
Gebhardt~\etal\ (2000a). Both galaxies were observed using the
0.21\arcsec\ square aperture. The wavelength range, 4566--6815~\AA,
includes the Mg~I~{\it b} lines near 5175~\AA. The spectral dispersion
is 1.09~\AA\ pixel$^{-1}$. The instrumental velocity dispersion is
$\sigma_{\rm instr} = $FWHM/2.35$ = 1.76 \pm 0.03$ pixels = $101\pm
2$~\kms\ (internal error). This width is intrinsic to the instrument
and is not strongly affected by how the aperture is illuminated
(Keyes~\etal\ 1995). We therefore make no aperture illumination
corrections to the measured velocity dispersion. Flat fielding and
correction for geomagnetically induced motions (GIM) were done as in
Kormendy~\etal\ (1996). The flat-field image uses the same
aperture. Each galaxy exposure is divided into multiple
subintegrations during the visits. There are four individual exposures
for NGC~5845 with a total integration time of 2.43 hours. For
NGC~3377, the central pointing had a single exposure of 0.66 hours,
and two flanking exposures on both sides of the galaxy of 1.02 and
0.90 hours (each split into two subintegrations).

The most important step is to determine where the slit was actually
placed.  For NGC~3379 (Gebhardt~\etal\ 2002a), this was a critical
issue since the aperture was not placed exactly in the center of the
galaxy. For both NGC~3377 and NGC~5845, the aperture placement is much
more secure since both galaxies have central cusps. Since we take a
setup image of the galaxy before the spectral observations, we know
where the aperture was placed. For both galaxies, the center of the
galaxy is at the center of the FOS aperture to better than
0.05\arcsec. For NGC~3377, we have two additional apertures placed
0.2\arcsec\ away from the center along the major axis on opposite
sides. We have checked their placement using the setup images and
confirm that they were both placed at the requested position to within
0.05\arcsec. Since the flanking spectra were placed at very similar
radii on opposite sides of the galaxy, we have fit the same velocity
profile, but appropriately flipped, to both spectra. This fit is the
same as that done for all of the other data used in the
models. Table~4, therefore, only reports the moments for one profile
fitted to both spectra.

We have three template stars taken with the same FOS aperture, and all
three provide similar results for the kinematics. Table~4 includes the
Gauss-Hermite moments for the single pointing for NGC~5845 and the two
pointings for NGC~3377.

Extraction of the LOSVD for the three {\it HST} spectra and the
ground-based spectra use the procedure described by Gebhardt~\etal\
(2000a) and Pinkney~\etal\ (2002a). We use the full LOSVD in the
models, but we report only its first four moments in Table~4.


\hskip 50pt\psfig{file=gebhardt.fig14.ps,width=12.0cm,angle=0}
\figcaption[gebhardt.fig14.ps]{The first four moments of the LOSVD for
NGC~5845 along three different position angles. The black circles are
along the major axis. The red squares are along the axis 20\degr\ up
from the major axis, and the blue triangles are along the minor
axis. The light-blue diamonds represent the first four moments of the
single FOS pointing that we have at the center.}


\begin{deluxetable}{lcccccccccc}
\footnotesize
\tablecolumns{11}
\tablewidth{0pt}
\tablenum{1}
\tablecaption{Galaxy sample}
\tablehead{
\colhead{(1)} &
\colhead{(2)} &
\colhead{(3)} &
\colhead{(4)} &
\colhead{(5)} &
\colhead{(6)} &
\colhead{(7)} &
\colhead{(8)} &
\colhead{(9)} &
\colhead{(10)} &
\colhead{(11)} \\
\colhead{Galaxy}                   & 
\colhead{Type}                     & 
\colhead{$M_B$}                    & 
\colhead{$M_{BH}$ (low,high)}   &
\colhead{$\sigma_e$}               & 
\colhead{Dist}                     & 
\colhead{$M/L$,}                   &
\colhead{$\bigr({{{\rm dlog}(\nu)}\over{{\rm dlog}(r)}}\bigl)_0$}&
\colhead{$\bigr({{\sigma_r}\over{\sigma_t}}\bigl)_0$}&
\colhead{$\bigr({{\sigma_r}\over{\sigma_t}}\bigl)_{{\rm R}_e/4}$}&
\colhead{$R_e$}           \\
\colhead{} & \colhead{} & \colhead{bulge} & \colhead{$M_\odot$} & \colhead{\kms} & 
\colhead{Mpc} & \colhead{band} & \colhead{} & \colhead{} & \colhead{} & \colhead{kpc} 
}
\startdata
N821      & E4  & $-$20.41 & $3.7\times 10^7~(2.9,6.1)$ & 209 &  24.1  & 7.6,$V$ & --1.4 & 0.37 & 0.88 & 5.32 \\
N2778     & E2  & $-$18.59 & $1.4\times 10^7~(0.5,2.2)$ & 175 &  22.9  & 8.0,$V$ & --1.9 & 0.56 & 0.81 & 1.82 \\ 
N3377     & E5  & $-$19.05 & $1.0\times 10^8~(0.9,1.9)$ & 145 &  11.2  & 2.9,$V$ & --1.5 & 0.74 & 1.01 & 1.82 \\
N3384     & S0  & $-$18.99 & $1.6\times 10^7~(1.4,1.7)$ & 143 &  11.6  & 2.5,$V$ & --1.9 & 0.44 & 0.89 & 0.73 \\
N3608$^*$ & E2  & $-$19.86 & $1.9\times 10^8~(1.3,2.9)$ & 182 &  22.9  & 3.7,$V$ & --1.0 & 0.52 & 1.00 & 3.85 \\
N4291$^*$ & E2  & $-$19.63 & $3.1\times 10^8~(0.8,3.9)$ & 242 &  26.2  & 5.5,$V$ & --0.6 & 0.42 & 0.98 & 1.85 \\
N4473$^*$ & E5  & $-$19.89 & $1.1\times 10^8~(0.3,1.5)$ & 190 &  15.7  & 6.0,$V$ & --0.3 & 0.33 & 0.79 & 1.84 \\
N4564     & E3  & $-$18.92 & $5.6\times 10^7~(4.8,5.9)$ & 162 &  15.0  & 2.0,$I$ & --1.9 & 0.62 & 0.89 & 1.54 \\
N4649$^*$ & E1  & $-$21.30 & $2.0\times 10^9~(1.4,2.4)$ & 385 &  16.8  & 8.5,$V$ & --1.2 & 0.51 & 1.04 & 5.95 \\
N4697     & E4  & $-$20.24 & $1.7\times 10^8~(1.6,1.9)$ & 177 &  11.7  & 4.7,$V$ & --1.7 & 0.97 & 0.92 & 4.25 \\
N5845     & E3  & $-$18.72 & $2.4\times 10^8~(1.0,2.8)$ & 234 &  25.9  & 5.5,$V$ & --1.4 & 0.79 & 1.07 & 0.51 \\
N7457     & S0  & $-$17.69 & $3.5\times 10^6~(2.1,4.6)$ &  67 &  13.2  & 3.2,$V$ & --1.9 & 0.62 & 0.72 & 0.90 \\
\enddata

\tablecomments{An asterisk denotes a core galaxy; the others are
power-law galaxies. The bulge, B-band magnitudes come from Kormendy \&
Gebhardt (2001). $\sigma_e$ comes from the ground-based spectra
integrating from $\pm$R$_e$ along the major axis with a 1\arcsec\
slit. The black hole mass offsets used in Fig~9 represent the
difference between the black hole mass in this table from that mass
using the correlation from Tremaine~\etal\ (2002) with the $\sigma_e$
reported in this table. The distances come from Tonry~\etal\
(2001). The mass-to-light ratios come from the best-fit for the
dynamical models presented in this paper, and generally have
uncertainties smaller than 2\%. The central radial to tangential
dispersion ratio come from an average of the value in central bins
along those position angles for which we have kinematic data (from one
to three position angles; see Table~2).  The ratio at R$_e$/4 is an
average of the three bins nearest in radii to R$_e$/4 along the
position angles that have kinematic data. R$_e$ comes from either
Faber~\etal\ (1989) or Baggett~\etal\ (2000).}
\end{deluxetable}

\begin{deluxetable}{lllclrr}
\tablewidth{0pt}
\tablenum{2}
\tablecaption{Model Parameters}
\tablehead{
\colhead{(1)} &
\colhead{(2)} &
\colhead{(3)} &
\colhead{(4)} &
\colhead{(5)} &
\colhead{(6)} &
\colhead{(7)} \\
\colhead{Galaxy}                   & 
\colhead{$M_{BH}$ (low,high)}      &
\colhead{$M_{BH}$ (low,high)}      &
\colhead{N$_{\rm PA}$}             &
\colhead{N$_{\rm Pos}$}            &
\colhead{N$_{\rm Fit}$}            &
\colhead{$\chi^2$}                 \\
\colhead{}                         & 
\colhead{$M_\odot$, All Data}      &
\colhead{$M_\odot$, Ground Only}   &
\colhead{({\it HST}, Ground)}      &
\colhead{}                         &
\colhead{}                         &
\colhead{Minimum}                  }
\startdata
N821      & $3.7\times10^7~(2.9,6.1)$ & $3.0\times10^7~(0.0,8.0)$ & 2 (1,2) & 24 & 312 & 128 \\
N2778     & $1.4\times10^7~(0.5,2.2)$ & $0.0\times10^7~(0.0,6.0)$ & 1 (1,1) & 12 &  93 &  29 \\
N3377     & $1.0\times10^8~(0.9,1.9)$ & $1.2\times10^8~(0.5,2.0)$ & 2 (1,2) & 15 &  52 &   8 \\
N3384     & $1.6\times10^7~(1.4,1.7)$ & $1.4\times10^7~(1.1,3.0)$ & 2 (1,2) & 24 & 312 & 131 \\
N3608     & $1.9\times10^8~(1.3,2.9)$ & $1.4\times10^8~(0.7,3.0)$ & 2 (1,2) & 24 & 312 &  92 \\
N4291     & $3.1\times10^8~(0.8,3.9)$ & $2.0\times10^8~(0.0,5.0)$ & 3 (1,3) & 27 & 351 & 182 \\
N4473     & $1.1\times10^8~(0.3,1.5)$ & $2.0\times10^7~(1.0,9.9)$ & 2 (1,2) & 24 & 312 &  64 \\
N4564     & $5.6\times10^7~(4.8,5.9)$ & $1.0\times10^7~(0.6,5.5)$ & 3 (1,3) & 33 & 429 & 187 \\
N4649     & $2.0\times10^9~(1.4,2.4)$ & $1.5\times10^9~(0.7,2.5)$ & 3 (1,3) & 35 & 455 & 128 \\
N4697     & $1.7\times10^8~(1.6,1.9)$ & $2.5\times10^8~(1.6,3.1)$ & 3 (1,3) & 30 & 390 & 192 \\
N5845     & $2.4\times10^8~(1.0,2.8)$ & $3.0\times10^8~(0.4,4.5)$ & 3 (1,3) & 25 & 325 & 205 \\
N7457     & $3.5\times10^6~(2.1,4.6)$ & $3.1\times10^6~(0.0,9.9)$ & 3 (1,3) & 20 & 260 &  76 \\
\enddata

\tablecomments{The BH masses come from fitting to all of the data
(col.~2) and fitting only to the ground-based spectral data
(col.~3). N$_{\rm PA}$ (col.~4) is the number of kinematic position
angles, with the number of {\it HST} and ground-based PAs given in
parenthesis. N$_{\rm Pos}$ (col.~5) is the total number of positions
on the sky with kinematics (this includes both ground-based and {\it
HST} data). N$_{\rm Fit}$ (col.~6) is the number of kinematic data
points used in the fits; for most galaxies we use 13 velocity bins to
represent the LOSVD.}

\end{deluxetable}

\begin{deluxetable}{lcccccc}
\tablewidth{0pc}
\tablenum{3}
\tablecaption{Galaxies from Other Sources}
\tablehead{
\colhead{Galaxy}                   & 
\colhead{$M_{BH}$ (low,high)}      &
\colhead{${{{\rm dlog}(\nu)}\over{{\rm dlog}(r)}}_0$}&
\colhead{$\bigr({{\sigma_r}\over{\sigma_t}}\bigl)_0$}&
\colhead{$\bigr({{\sigma_r}\over{\sigma_t}}\bigl)_{{\rm R}_e/4}$}&
\colhead{$R_e$ (kpc)}     &
\colhead{Ref}                      }
\startdata
N221=M32  & $2.5\times10^6~(2.4,2.6)$ & --1.6 & 1.01 & 0.72 & 0.15 & 1\\
N1023     & $4.4\times10^7~(3.9,4.9)$ & --1.8 & 0.57 & 1.04 & 1.98 & 2\\
N3379$^*$ & $1.0\times10^8~(0.5,1.6)$ & --1.0 & 0.41 & 1.06 & 1.76 & 3\\
N4342     & $3.0\times10^8~(2.4,4.1)$ & --1.7 & 1.03 & 1.00 & 0.47 & 4\\
IC1459    & $2.5\times10^9~(2.4,2.6)$ & --1.4 & 0.81 & 1.12 & 4.48 & 5\\
\enddata

\tablecomments{References: (1) Verolme~\etal\ (2002); (2) Bower~\etal\
(2001); (3) Gebhardt~\etal\ (2000); (4) Cretton \& van den Bosch
(1999); (5) Cappellari~\etal\ (2002). An asterisk denotes a core
galaxy. The uncertainties on the BH masses are 68\% for one degree of
freedom. For M32, N4342, and IC1459 we approximate these $1\sigma$
uncertainties from the authors quoted uncertainties.}

\end{deluxetable}

\newcommand{\spa}{\phantom{1}}

\begin{deluxetable}{rrrrrr}
\tablecolumns{6}
\tablewidth{0pt}
\tablenum{4}
\tablecaption{Kinematic data for NGC~3377 and NGC~5845}
\tablehead{
\colhead{PA}                   & 
\colhead{Radius (\arcsec)}     & 
\colhead{Velocity (\kms)}      & 
\colhead{$\sigma$ (\kms)}      & 
\colhead{H3}                   & 
\colhead{H4}                   }
\startdata
\multicolumn{6}{c}{{\it HST}/FOS Kinematics for NGC~3377} \nl
 0\degr &  0.00 &    0.0$\pm$\spa6.5 &  258.0$\pm$\spa6.0 &--0.05$\pm$0.02 &  0.00$\pm$0.02 \\
 0\degr &--0.20 &  100.0$\pm$\spa5.0 &  215.0$\pm$\spa5.0 &--0.06$\pm$0.02 &--0.01$\pm$0.02 \\
\tablevspace{1pt}
\multicolumn{6}{c}{{\it HST}/FOS Kinematics for NGC~5845} \nl
 0\degr &  0.00 &   17.2$\pm$20.8    &  292.9$\pm$17.0    &--0.09$\pm$0.05 &--0.05$\pm$0.04 \\
\tablevspace{1pt}
\multicolumn{6}{c}{Ground-Based Kinematics for NGC~5845} \nl
 0\degr &--0.13 &   63.0$\pm$\spa7.8 &  250.0$\pm$\spa7.2 &  0.03$\pm$0.03 &--0.02$\pm$0.02 \\
 0\degr &  0.46 & --36.6$\pm$11.5    &  239.4$\pm$\spa8.3 &  0.10$\pm$0.02 &  0.02$\pm$0.03 \\
 0\degr &  1.05 & --76.7$\pm$12.2    &  224.9$\pm$13.3    &  0.11$\pm$0.03 &--0.01$\pm$0.04 \\
 0\degr &  1.93 &--122.6$\pm$\spa7.3 &  185.2$\pm$\spa6.3 &  0.07$\pm$0.02 &--0.02$\pm$0.02 \\
 0\degr &  3.41 & --95.8$\pm$10.1    &  170.3$\pm$\spa6.7 &--0.02$\pm$0.04 &--0.09$\pm$0.01 \\
 0\degr &  5.77 & --44.9$\pm$12.3    &  140.4$\pm$18.1    &  0.09$\pm$0.09 &  0.01$\pm$0.07 \\
 0\degr &  9.60 &--103.0$\pm$27.3    &  170.7$\pm$33.5    &  0.15$\pm$0.09 &  0.11$\pm$0.08 \\
 0\degr & 15.80 & --21.3$\pm$53.3    &  163.5$\pm$68.9    &  0.10$\pm$0.12 &  0.01$\pm$0.20 \\
 0\degr &--0.72 &  138.7$\pm$\spa9.4 &  226.2$\pm$10.4    &--0.09$\pm$0.04 &--0.00$\pm$0.03 \\
 0\degr &--1.31 &  173.4$\pm$\spa9.8 &  191.2$\pm$12.9    &--0.14$\pm$0.04 &  0.03$\pm$0.03 \\
 0\degr &--1.90 &  172.5$\pm$\spa9.1 &  194.5$\pm$11.1    &--0.03$\pm$0.04 &--0.04$\pm$0.03 \\
 0\degr &--2.79 &  143.1$\pm$\spa7.9 &  175.7$\pm$\spa7.6 &  0.01$\pm$0.03 &--0.05$\pm$0.03 \\
 0\degr &--4.26 &   92.6$\pm$\spa6.2 &  138.4$\pm$10.3    &--0.05$\pm$0.03 &--0.05$\pm$0.02 \\
 0\degr &--6.62 &  142.9$\pm$17.1    &  161.6$\pm$27.0    &--0.07$\pm$0.07 &  0.00$\pm$0.07 \\
 0\degr &--10.16&  171.7$\pm$20.2    &  154.6$\pm$24.6    &--0.01$\pm$0.07 &--0.04$\pm$0.06 \\
22\degr &--0.33 &   34.7$\pm$11.4    &  253.0$\pm$\spa9.8 &  0.01$\pm$0.03 &--0.01$\pm$0.04 \\
22\degr &  0.26 & --45.8$\pm$\spa8.1 &  230.4$\pm$\spa8.1 &  0.07$\pm$0.02 &--0.04$\pm$0.03 \\
22\degr &  0.85 &--102.8$\pm$\spa9.6 &  209.6$\pm$\spa9.1 &  0.11$\pm$0.03 &  0.00$\pm$0.03 \\
22\degr &  1.74 &--105.7$\pm$\spa7.8 &  197.7$\pm$\spa7.2 &  0.11$\pm$0.02 &--0.01$\pm$0.02 \\
22\degr &  3.21 & --71.0$\pm$\spa8.5 &  175.0$\pm$\spa8.1 &  0.04$\pm$0.03 &  0.02$\pm$0.03 \\
22\degr &  5.57 & --59.4$\pm$10.2    &  147.2$\pm$10.4    &--0.05$\pm$0.03 &--0.02$\pm$0.03 \\
22\degr &  9.41 & --52.2$\pm$20.9    &  142.0$\pm$43.1    &  0.04$\pm$0.06 &--0.03$\pm$0.16 \\
22\degr &--0.92 &   93.6$\pm$10.9    &  237.2$\pm$10.9    &--0.05$\pm$0.04 &  0.02$\pm$0.04 \\
22\degr &--1.51 &  119.3$\pm$\spa8.7 &  203.0$\pm$\spa7.6 &--0.04$\pm$0.03 &--0.05$\pm$0.03 \\
22\degr &--2.39 &  113.9$\pm$\spa7.6 &  189.0$\pm$\spa5.8 &--0.00$\pm$0.03 &--0.05$\pm$0.03 \\
22\degr &--3.87 &  112.1$\pm$\spa7.2 &  162.4$\pm$\spa6.4 &  0.03$\pm$0.03 &--0.03$\pm$0.02 \\
22\degr &--6.23 &   87.8$\pm$\spa9.7 &  139.4$\pm$14.5    &  0.00$\pm$0.05 &--0.03$\pm$0.04 \\
22\degr &--10.06&  139.1$\pm$29.7    &  160.7$\pm$31.7    &--0.18$\pm$0.12 &  0.11$\pm$0.15 \\
90\degr &  0.07 &   42.6$\pm$11.7    &  280.1$\pm$11.4    &  0.05$\pm$0.04 &  0.02$\pm$0.04 \\
90\degr &  0.66 &   49.8$\pm$11.7    &  246.6$\pm$\spa8.6 &--0.03$\pm$0.04 &--0.07$\pm$0.03 \\
90\degr &  1.25 &   65.5$\pm$\spa8.7 &  218.8$\pm$\spa6.6 &--0.01$\pm$0.02 &--0.01$\pm$0.03 \\
90\degr &  2.13 &   31.5$\pm$\spa8.3 &  173.5$\pm$\spa7.5 &--0.07$\pm$0.03 &--0.05$\pm$0.02 \\
90\degr &  3.61 &   42.0$\pm$\spa6.1 &  159.5$\pm$\spa8.6 &  0.00$\pm$0.02 &--0.05$\pm$0.01 \\
90\degr &  5.67 &   54.3$\pm$14.6    &  133.6$\pm$13.2    &--0.00$\pm$0.04 &--0.04$\pm$0.04 \\
90\degr &  9.21 &   21.9$\pm$52.0    &  160.0$\pm$75.3    &  0.01$\pm$0.25 &  0.01$\pm$0.25 \\
90\degr &--0.52 &   38.5$\pm$12.0    &  236.5$\pm$11.0    &  0.05$\pm$0.04 &--0.01$\pm$0.03 \\
90\degr &--1.11 &   20.6$\pm$10.9    &  223.8$\pm$\spa6.9 &  0.06$\pm$0.04 &--0.04$\pm$0.03 \\
90\degr &--1.70 &   29.2$\pm$11.0    &  197.3$\pm$10.4    &  0.04$\pm$0.03 &--0.05$\pm$0.03 \\
90\degr &--2.59 &   20.2$\pm$12.3    &  183.9$\pm$13.6    &  0.06$\pm$0.06 &--0.02$\pm$0.02 \\
90\degr &--4.06 &   27.1$\pm$10.3    &  155.4$\pm$\spa8.7 &  0.01$\pm$0.03 &--0.03$\pm$0.02 \\
90\degr &--6.42 &   35.9$\pm$19.3    &  167.6$\pm$18.3    &  0.10$\pm$0.07 &--0.01$\pm$0.05 \\
90\degr &--10.26&   78.5$\pm$53.7    &  195.8$\pm$93.1    &  0.07$\pm$0.16 &--0.03$\pm$0.27 \\

\enddata

\end{deluxetable}

\end{document}